 \documentclass[12pt,preprint]{aastex}

\usepackage{epsfig,graphicx,natbib}
\usepackage{amssymb}
\usepackage{amsfonts}
\usepackage{amsmath}
\usepackage{color}
\usepackage{lscape}

\shorttitle{Molecular Emission Trends from Extragalactic IMFs} \shortauthors{Banerji et al.}

\begin{document}


\title{Trends in Molecular Emission from Different Extragalactic Stellar Initial Mass Functions}


\author{M. Banerji\altaffilmark{1},
  S. Viti\altaffilmark{1},
  D.A. Williams\altaffilmark{1}}

\email{mbanerji@star.ucl.ac.uk}


\altaffiltext{1}{Department of Physics and Astronomy, University
  College London, Gower Street, London WC1E 6BT, UK.}


\begin{abstract}
Banerji et al. (2009) suggested that top-heavy stellar Initial Mass Functions (IMFs) in galaxies may arise 
when the interstellar physical conditions inhibit low-mass star formation, 
and they determined the physical conditions under which this suppression 
may or may not occur. In this work, we explore the sensitivity of 
the chemistry of interstellar gas under a wide range of conditions. We use 
these results to predict the relative velocity-integrated antenna 
temperatures of the CO rotational spectrum for several models of high 
redshift active galaxies which may produce both top-heavy and unbiased IMFs. We find that while active galaxies with solar 
metallicity (and top-heavy IMFs) produce higher antenna temperatures than 
those with sub-solar metallicity (and unbiased IMFs) the actual 
rotational distribution is similar. The high-J to peak CO ratio however may be used to roughly infer the metallicity of a galaxy provided we know whether it is active or quiescent. The metallicity strongly influences the shape of the IMF. High order CO transitions are also found to
provide a good diagnostic for high far-UV intensity and low metallicity counterparts of Milky Way type systems both of which show some evidence for having top-heavy IMFs. We also compute 
the relative abundances of molecules known to be effective tracers of high 
density gas in these galaxy models. We find that the molecules CO and CS may be used to distinguish between solar and sub-solar metallicity in active galaxies at high redshift whereas HCN, HNC and CN are found to be relatively insensitive to the IMF shape at the large visual magnitudes typically associated with extragalactic sources.

\end{abstract}

\keywords{astrochemistry-galaxies:active - ISM:abundances - ISM:molecules - stars:formation}

\section{Introduction}

In \citet{Banerji:timescales} (B09 hereafter) we investigated timescales relevant for low-mass star formation for galaxies with a variety of different physical conditions. That study allowed us to infer in a crude way the physical properties of galaxies in which low-mass star formation is likely to be impeded possibly resulting in a top-heavy Initial Mass Function (IMF). We concluded that low-mass star formation is likely to be inhibited in low-metallicity systems as well as systems with high cosmic ray ionisation rates and FUV radiation field strengths where the magnetic pressure dominates over the thermal pressure. Such systems may therefore have a high-mass biased IMF. We also concluded that active galaxies at high redshift with sub-solar metallicities show enhanced low-mass star formation resulting in an unbiased IMF. However, similarly active galaxies with solar metallicities may have low-mass star formation impeded by magnetic pressure resulting in a high-mass biased IMF. 

The aim of this follow-up paper is to now consider the interstellar chemistry associated with extragalactic regions with a variety of different physical conditions. By looking at the chemistry associated with the different extragalactic systems studied in B09, we can identify possible molecular tracers of different types of IMF that can be used to direct observations with future projects such as ALMA. In particular, we consider the rotational transitions of $^{12}$CO as well as molecular abundances associated with different physical conditions. In $\S$ \ref{sec:model} we describe our model for interstellar chemistry. In $\S$ \ref{sec:CO} we consider changes in the CO antenna temperature for the various rotational transitions for various physical conditions in the collapsing cloud. In $\S$ \ref{sec:abundances} we look at changes in the molecular abundances for different physical conditions and as a function of depth within the cloud. Finally, we summarise our results in $\S$ \ref{sec:disc}. 

\section{The Model}

\label{sec:model}

All results in this paper are obtained by modelling the gas-phase chemistry inside a molecular cloud using the UCL\_PDR code \citep{Bell:Xfactor1, Bell:Xfactor2} for photon-dominated regions which has already been benchmarked against several other PDR models \citep{Rollig:PDR}. The model is both time and depth dependent but has been run here at steady state corresponding to a time of 1Gyr. The cloud is treated as a semi-infinite 1D slab illuminated by an unidirectional flux of far ultraviolet photons. The code solves simultaneously the chemistry, thermal balance and radiative transfer within the cloud. The chemical network considered here has 83 species that interact in over 1300 reactions whose rates are taken from the UMIST database \citep{LeTeuff:UMIST99}. 

In the models considered in this work, the hydrogen atom number density, $n$, metallicity, $\xi$, cosmic ray ionisation rate, $\zeta$ and FUV radiation field strength, $\chi$ are the input parameters that are varied. The turbulent velocity of the cloud is fixed at 1.5 km s$^{-1}$ and the dust to gas mass ratio is assumed to scale linearly with metallicity. All the models used in this paper have already been presented in B09. 

Central to much of the work presented in this paper are the model predictions for parameters associated with the rotational lines of CO and we therefore describe how these are obtained in more detail. The gas in PDRs mainly cools through emissions from collisionally excited atoms and molecules as well as through interactions with cooler dust grains. The code calculates the emission from the [OI], [OII] and [CII] fine structure lines as well as the CO rotational lines at each depth and time step. This is done using the escape probability formalism of \citet{deJong:80} to solve the radiative transfer equations. Non local thermodynamic equilibrium (non-LTE) level populations are determined by solving the equations of statistical equilibrium using rate coefficients for CO collisions with H$_2$, H and He from \citet{Flower:85}, \citet{Green:76} and \citet{Green:78} respectively. The solutions of the radiative transfer equations give the emissivity, $\Lambda$, the intensity, $I$ and the opacity, $\tau$ for every line at every depth and time step. Several of these outputs are used throughout this paper. For further details of the UCL\_PDR code and the models considered in this work, please refer to B09 and references therein. 

Although this paper makes references to models of extragalactic systems, we emphasise that we are modelling the various PDR components that make up external galaxies rather than the galaxy as a whole. Our chosen input parameters therefore mimic the average conditions in individual PDRs within a galaxy. We choose to consider outputs at two visual extinctions of $A_v \sim3$ and $A_v \sim8$. These roughly correspond to the translucent and dense PDR gas components within a galaxy. Atomic and molecular emissions from different components within the galaxy will contribute independently to the line emissions detected from the galaxy as a whole and studied in detail in this paper.   Also, due to the fact that the PDR is modelled as a semi-infinite slab, a given visual extinction will correspond to different physical distances within the cloud under different physical conditions as the $A_v$ depends on both the number density and the metallicity of the cloud in our models. We come back to this point later in the paper.

\section{$^{12}$CO Rotational Transitions}

\label{sec:CO}

Carbon monoxide is one of the most commonly used tracers of molecular gas and its various rotational J transitions have been detected in many extragalactic systems both nearby \citep{Bradford:n253cr,Bayet:04,Bayet:06} and in the distant Universe \citep{Bertoldi:03,Nesvadba:COhighz,Daddi:COhighz,Knudsen:COhighz}. 
In Milky Way type conditions, the column density of molecular hydrogen along a line of sight is often related to the velocity integrated antenna temperature, T$_A$ for the CO(1-0) transition, using the CO-to-H$_2$ conversion factor or X factor \citep{Strong:96, Dame:01}. The CO(1-0) line is used as it is known to be the dominant CO cooling line with the highest antenna temperature under physical conditions appropriate for a Milky Way cloud. However, this is not necessarily the case for high redshift systems with significantly different physical conditions and higher order transitions of CO may become more relevant \citep{Sakamoto:99, Bell:Xfactor1, Bell:Xfactor2}. It is already known that galaxies with elevated cosmic ray fluxes have larger high-J CO line ratios \citep{Meijerink:cr} and that the higher order CO transitions trace denser gas than their lower order counterparts. In this section, we examine CO rotational line transitions and ratios in the context of the stellar initial mass function. We consider which of the CO transitions, if any, can be used as diagnostics for regions of both enhanced and inhibited low-mass star formation as studied in B09. We conduct this study both for systems where we vary one physical parameter at a time while keeping the others fixed at their local values as well as the three high redshift systems modelled in B09 whose input parameters are summarised in Table \ref{tab:highz} along with the IMFs associated with them and inferred in B09.

The UCL\_PDR code outputs the integrated line intensity of the CO rotational lines with 1 $\leq$ J$_{upper}$ $\leq$ 9 in units of erg s$^{-1}$ cm$^{-2}$ sr$^{-1}$ after solving the radiative transfer equations. The intensity is related to the emissivity $\Lambda$ by

\begin{equation}
I=\frac{1}{2\pi}\int \Lambda(L) dL
\label{eq:I}
\end{equation}

\noindent where $L$ is the depth into the cloud and the factor of $1/2 \pi$ takes into account the fact that the photons only emerge from the outer edge of the cloud due to the cloud being modelled as a semi-infinite slab. Integrating over larger depths will therefore obviously result in higher intensities.  

\begin{figure}[p]
\begin{center}
\begin{tabular}{cc}
\includegraphics[width=9cm,angle=0]{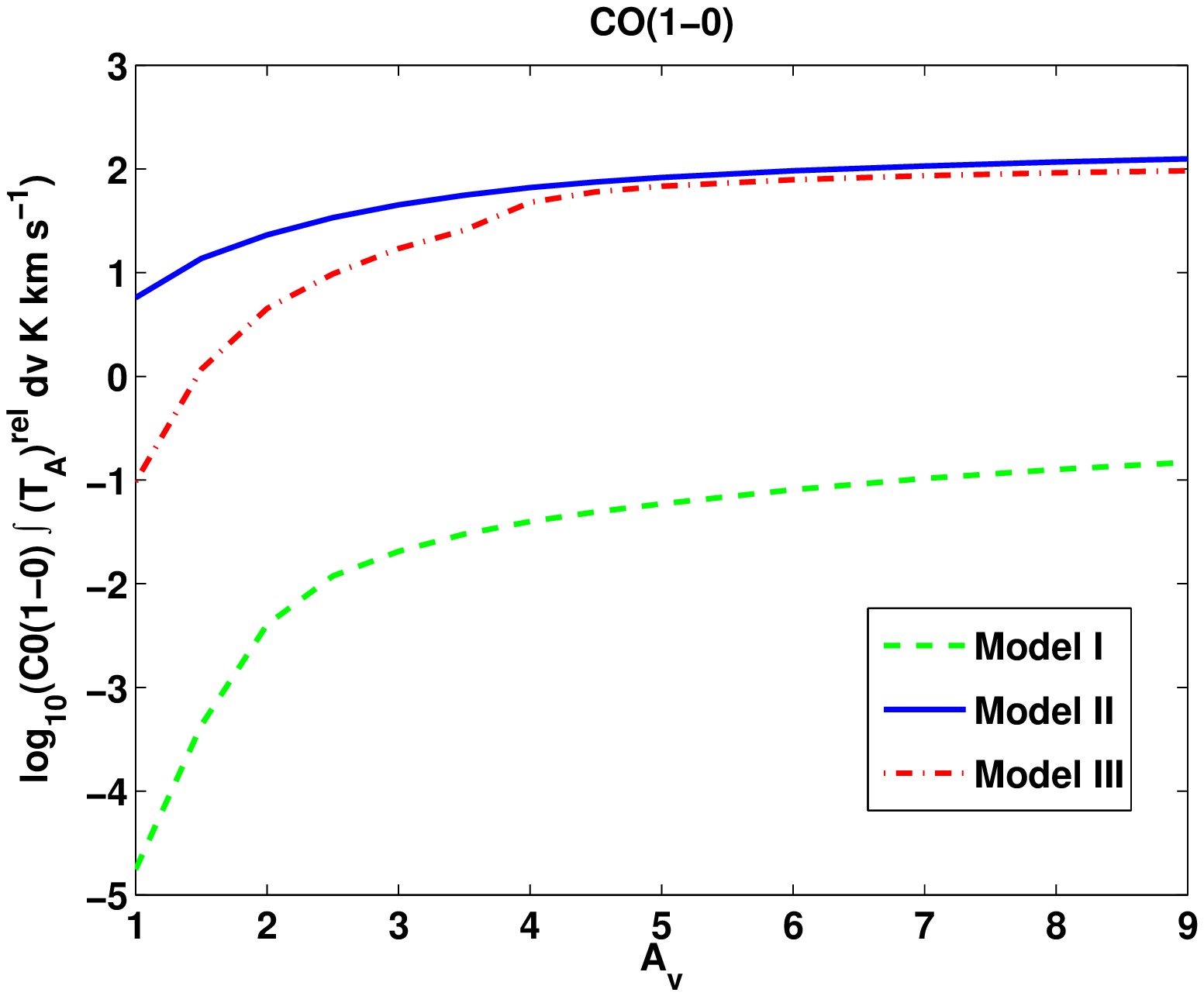} & \includegraphics[width=9cm,angle=0]{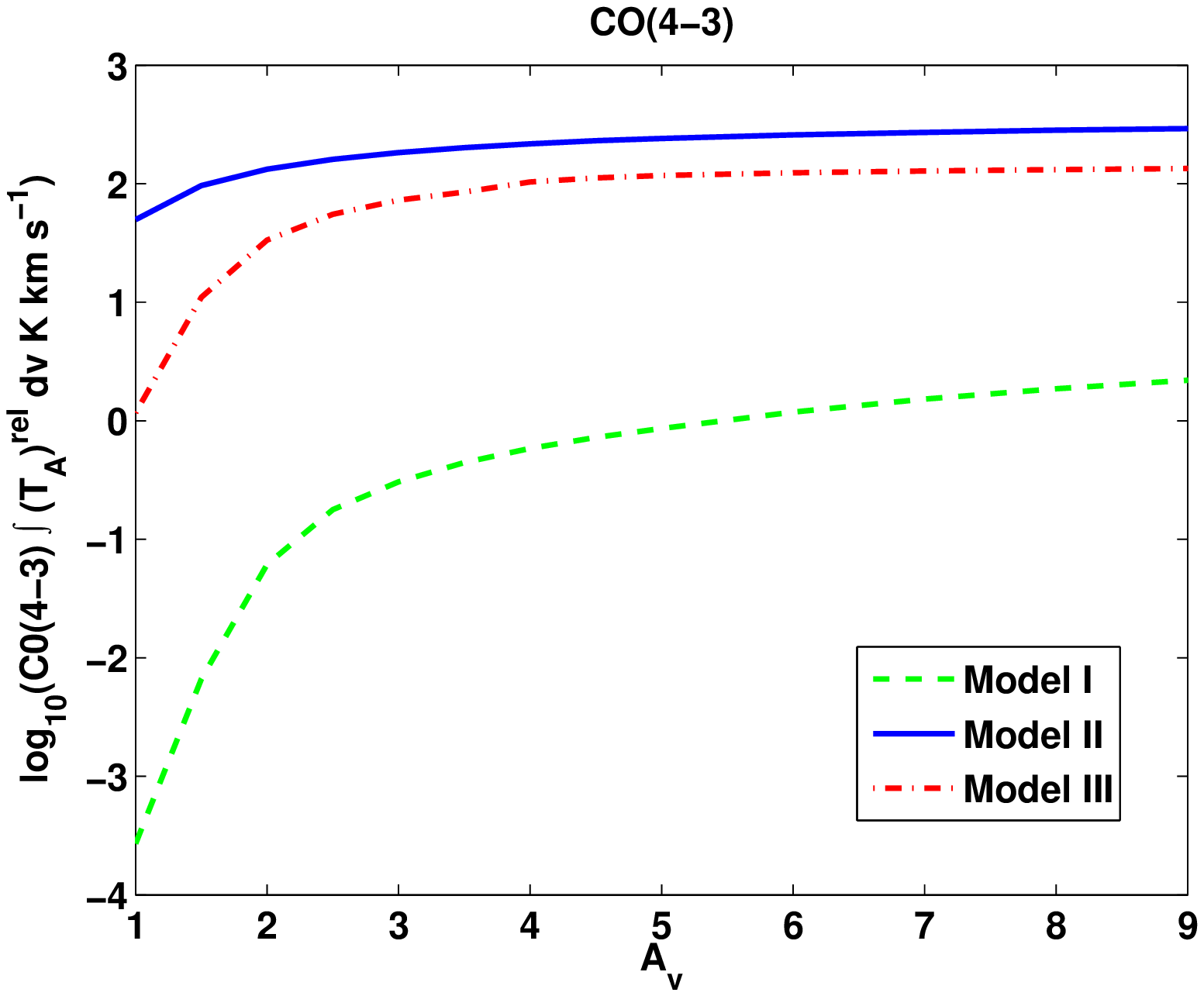} \\
\includegraphics[width=9cm,angle=0]{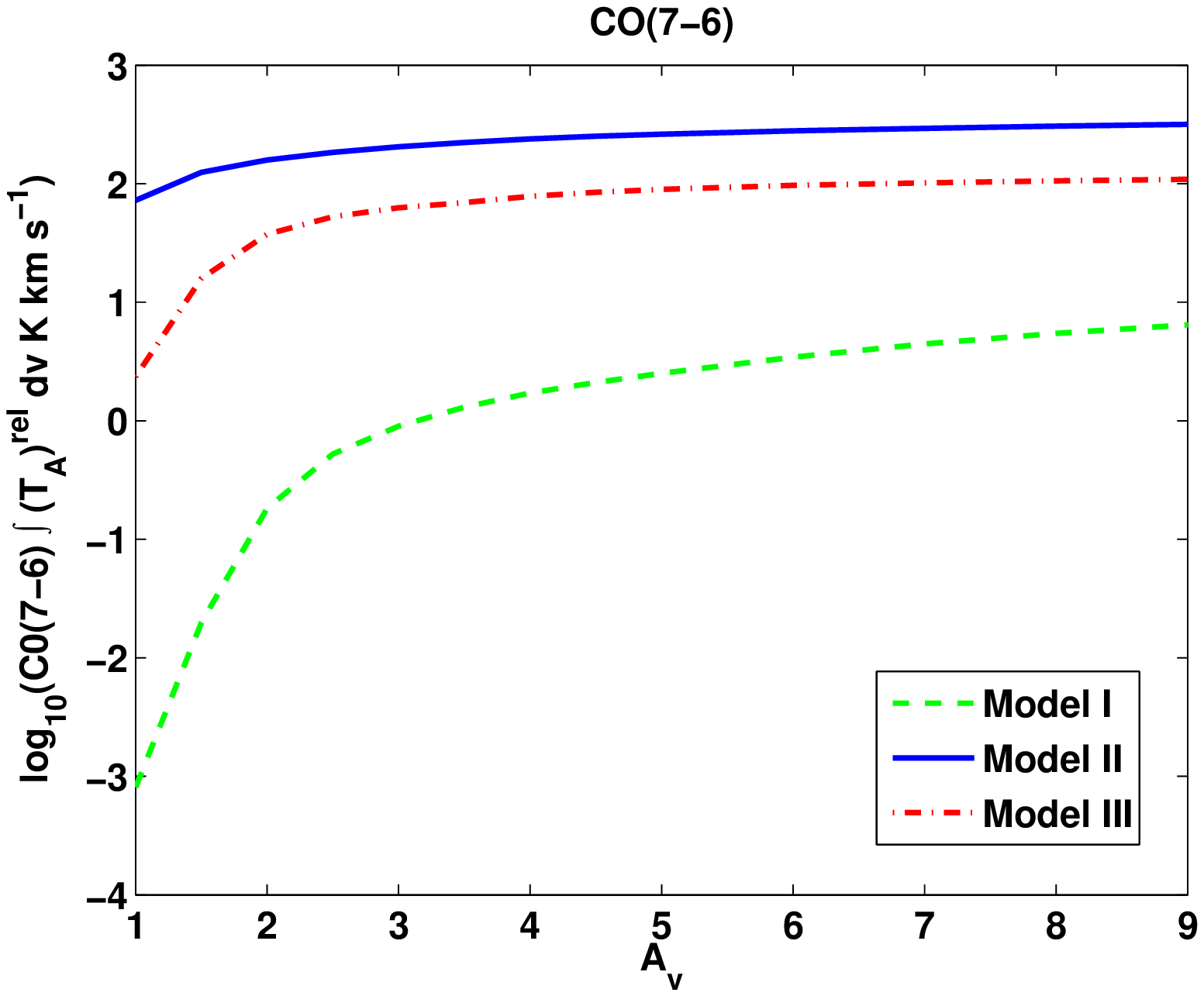} & \includegraphics[width=9cm,angle=0]{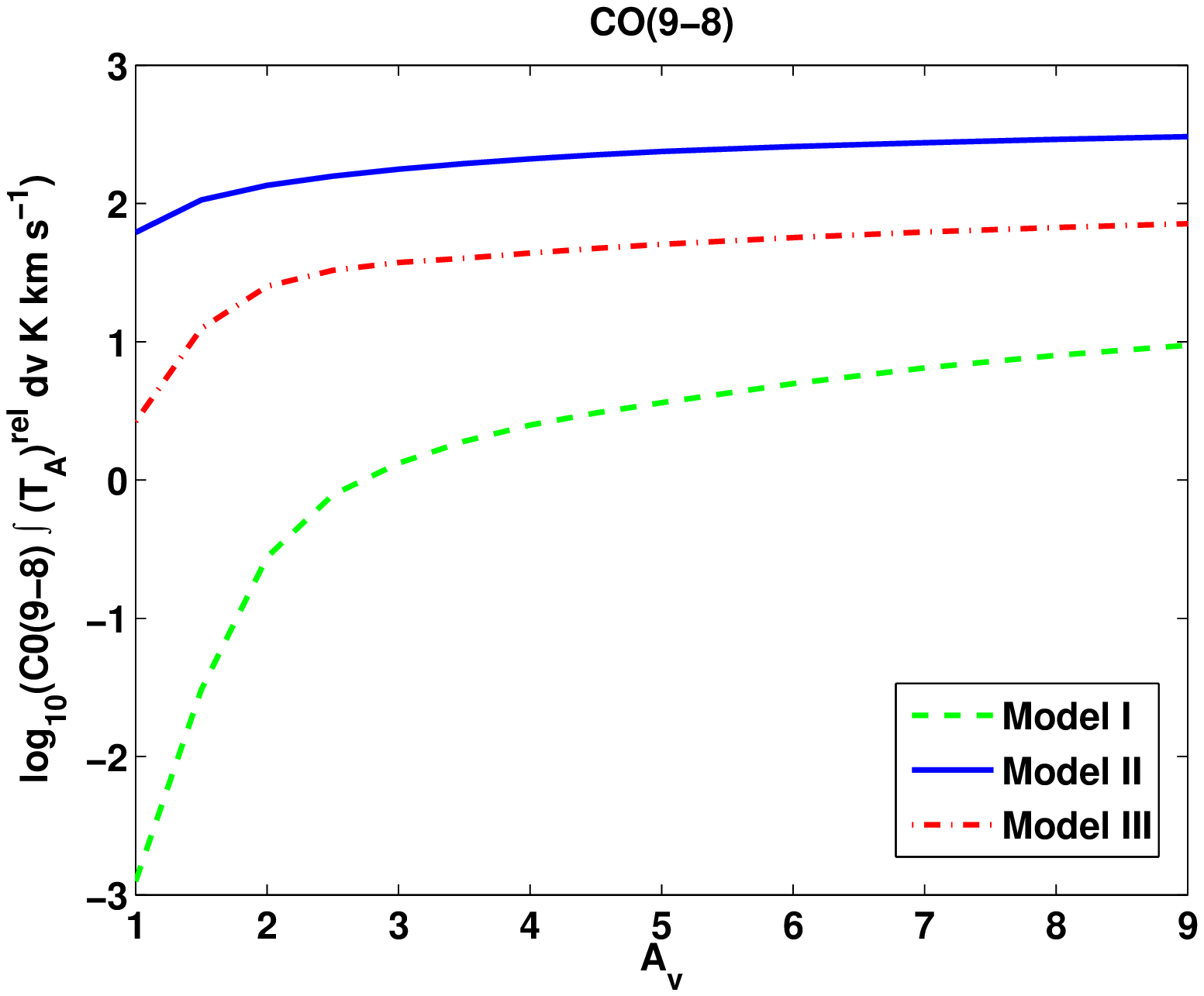} \\
\end{tabular}
\caption{Variation of theoretical velocity integrated CO antenna temperatures with $A_v$ for high redshift models (Table \ref{tab:highz}).}
\label{fig:IMF2}
\end{center}
\end{figure}

In order to calculate the relative velocity integrated antenna temparature from the intensity the following equation is used:

\begin{equation}
\int (T_A)^{rel} dv = 10^{-6}\frac{c^3I}{2k_B\nu^3 L_{A_v}} K km s^{-1} pc^{-1}
\label{eq:TA}
\end{equation}

\noindent where $c$ is the speed of light and $k_B$ is the Boltzmann constant. As the UCL\_PDR code models the PDR as a semi-infinite slab and the visual extinction, $A_v$ in the slab is related to both the number density as well as the metallicity, a fixed $A_v$ will correspond to different physical distances, $L$, for different values of the density and metallicity. In order to derive values for the relative temperatures that are comparable and independent of the size of the emitting region, we also divide the relative velocity integrated antenna temperatures by the distance in parsec corresponding to the $A_v$ at which the temperature is calculated, $L_{A_v}$. This distance will obviously be much larger for low density and low metallicity systems for a fixed $A_v$. Our figures therefore represent the relative velocity integrated antenna temperatures for the different $^{12}$CO transitions per unit parsec except in Figure \ref{fig:IMF2} where we plot the relative velocity integrated temperatures as a function of $A_v$. 

Note that the numbers calculated using Eq \ref{eq:TA} are important only in terms of observing trends. In order to match the absolute values of the theoretical velocity integrated antenna temperature to those from observations, one has to consider several other factors. Firstly, all models in B09 assume a turbulent velocity of 1.5 km s$^{-1}$ typical for a giant molecular cloud in the Galaxy. In reality, observed lines of CO in extragalactic sources typically have widths of several 100 km s$^{-1}$ due to contributions from several PDR regions within the galaxy and the temperature has to be scaled accordingly. Secondly, we need to account for a surface filling factor which takes into account the size of the source as well as the telescope beam. Finally, all observational results will be affected by factors such as the atmospheric conditions and the telescope efficiency. One way of comparing our theoretical predictions to observations however, is to consider line ratios rather than the intensities and brightness temperatures of individual lines. Assuming the emission from both lines comes from the same clouds, the various factors discussed above should cancel out when computing the ratio of either the intensity or the integrated temperature. We therefore compare our theoretical predictions to observed line ratios in $\S$ \ref{sec:obs}.  



In the following sections, we list in all cases our relative integrated temperatures per unit distance. We emphasise that the results presented here are only useful in terms of observing trends and should not be compared to absolute values obtained through observation as they do not take into account the factors already stated above.

\subsection{Density}

\label{sec:n}

\begin{figure}
\plotone{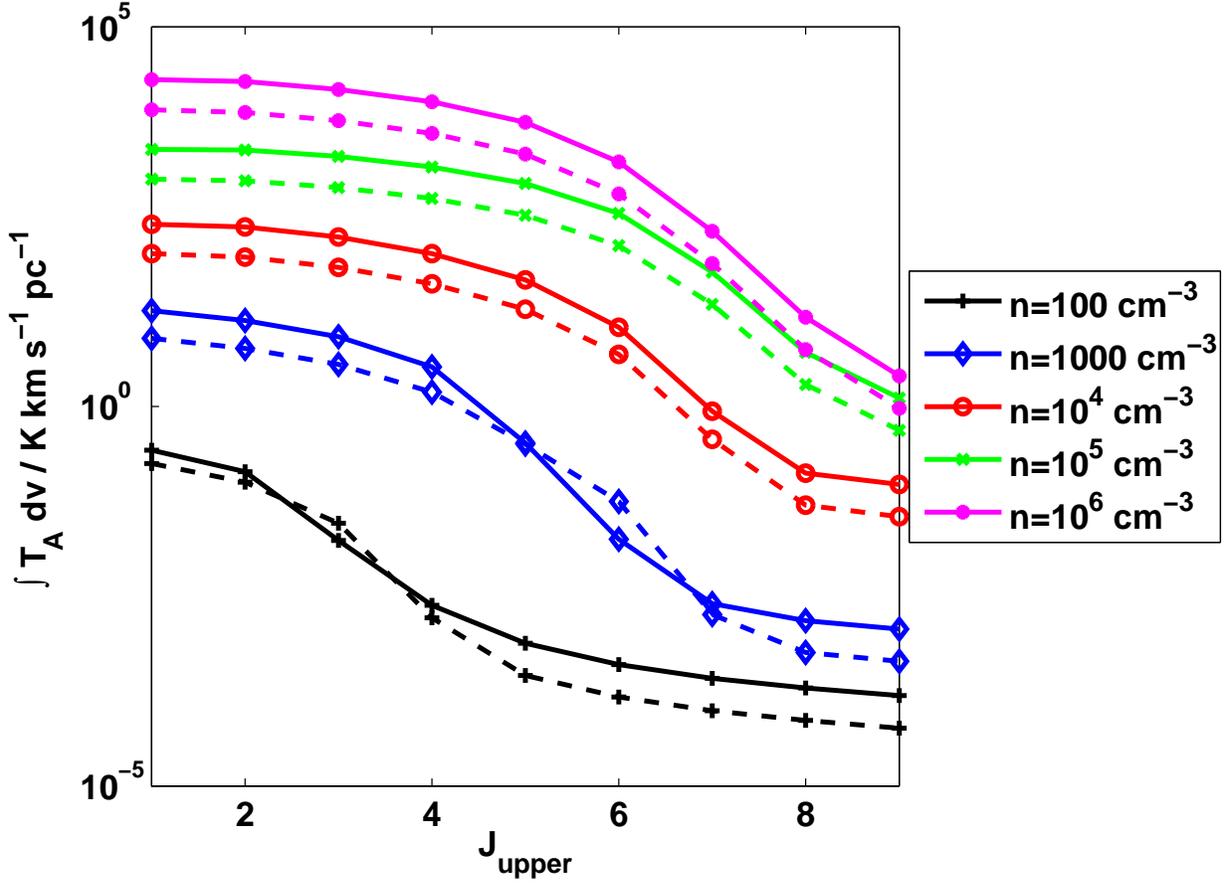}
\caption{Variation of relative velocity integrated CO antenna temperature per unit parsec with number density of hydrogen at $A_v\sim$3 (solid lines) and $A_v \sim 8$ (dashed lines). The metallicity, cosmic ray ionisation rate and incident FUV radiation intensity are fixed at $\xi=\xi_\odot$, $\zeta=10^{-17}s^{-1}$ and $\chi=1.7$.}
\label{fig:n}
\end{figure}

We vary the hydrogen atom number density of the molecular cloud between 100 cm$^{-3}$ and 10$^6$ cm$^{-3}$ while keeping the metallicity, cosmic ray ionisation rate and FUV flux fixed at $\xi=\xi_\odot$, $\zeta$=10$^{-17}$ s$^{-1}$ and $\chi$=1.7 Draine \citep{Draine:ISRF}. The effect on the relative velocity integrated antenna temperature, $\int (T_A)^{rel} dv$, of the different CO rotational transitions, is illustrated in Figure \ref{fig:n}. The CO(1-0) line is seen to be the brightest line at all densities and the CO SED falls off with increasing $J$. Higher density clouds also produce brighter antenna temperatures as expected. As the $A_v$ is increased, the relative integrated temperatures per unit parsec drop, particularly for the higher density clouds. This is telling us that in these high density clouds, most of the emissivity in these CO rotational lines is coming from the lower $A_v$ regions. This is also the case for the high-J transitions in the lower density clouds.  

\subsection{Metallicity}

\label{sec:xi}

\begin{figure}
\plotone{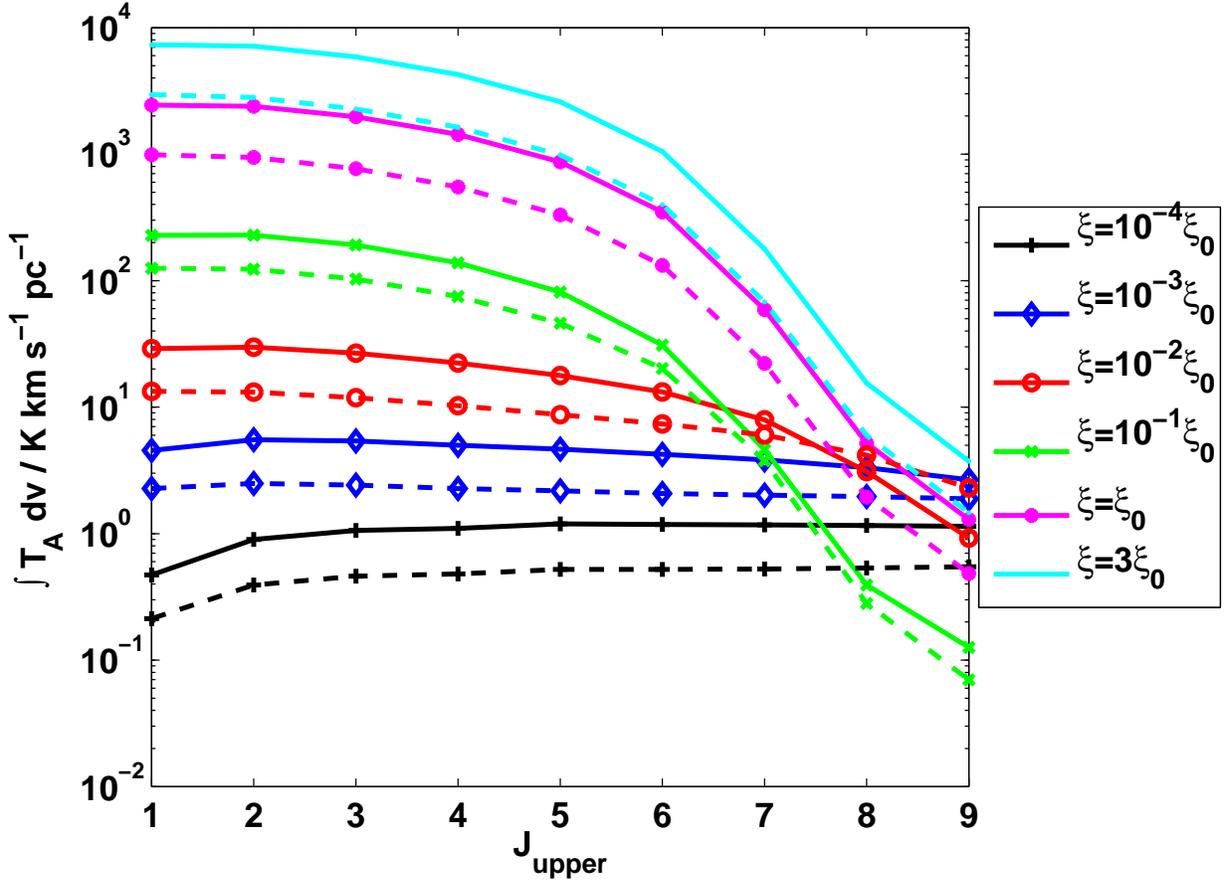}
\caption{Variation of relative velocity integrated CO antenna temperatures per unit parsec with metallicity at $A_v\sim$3 (solid lines) and $A_v \sim 8$ (dashed lines). The number density of hydrogen, cosmic ray ionisation rate and incident FUV radiation intensity are fixed at $n=10^5cm^{-3}$, $\zeta=10^{-17}s^{-1}$ and $\chi=1.7$.}
\label{fig:xi}
\end{figure}

We vary the metallicity between $\xi=10^{-4}\xi_\odot$ and $\xi=3\xi_\odot$ while keeping the density,  cosmic ray ionisation rate and FUV flux fixed at $n$=10$^{5}$ cm$^{-3}$, $\zeta$=10$^{-17}$ s$^{-1}$ and $\chi$=1.7 Draine. The dependence of $\int (T_A)^{rel} dv$ on the metallicity is illustrated in Figure \ref{fig:xi}. In B09 it was found that low-mass star formation is likely to be inhibited at very low metallicities resulting in a top-heavy IMF, provided the cloud is at moderately high densities and low cosmic ray fluxes and FUV field strengths. The form of the CO SED is quite different at different metallicities. While the higher metallicity clouds produce SEDs that peak at the lower rotational numbers and drop off sharply after that, the lower metallicity clouds have flatter SEDs. For the highest rotational transitions with $J_{upper}>7$, the sub-solar metallicity clouds actually produce emission that is brighter than that from solar metallicity clouds. As we approach sub-solar metallicities, the ratio of the high-order CO transitions to the peak CO transition tends towards unity whereas at solar metallicities it is much smaller. The physical conditions in sub-solar metallicity counterparts of Milky Way type galaxies are thought to be appropriate for the production of top-heavy IMFs.

\subsection{Cosmic-Ray Ionisation Rate}

\label{sec:cr}

\begin{figure}
\plotone{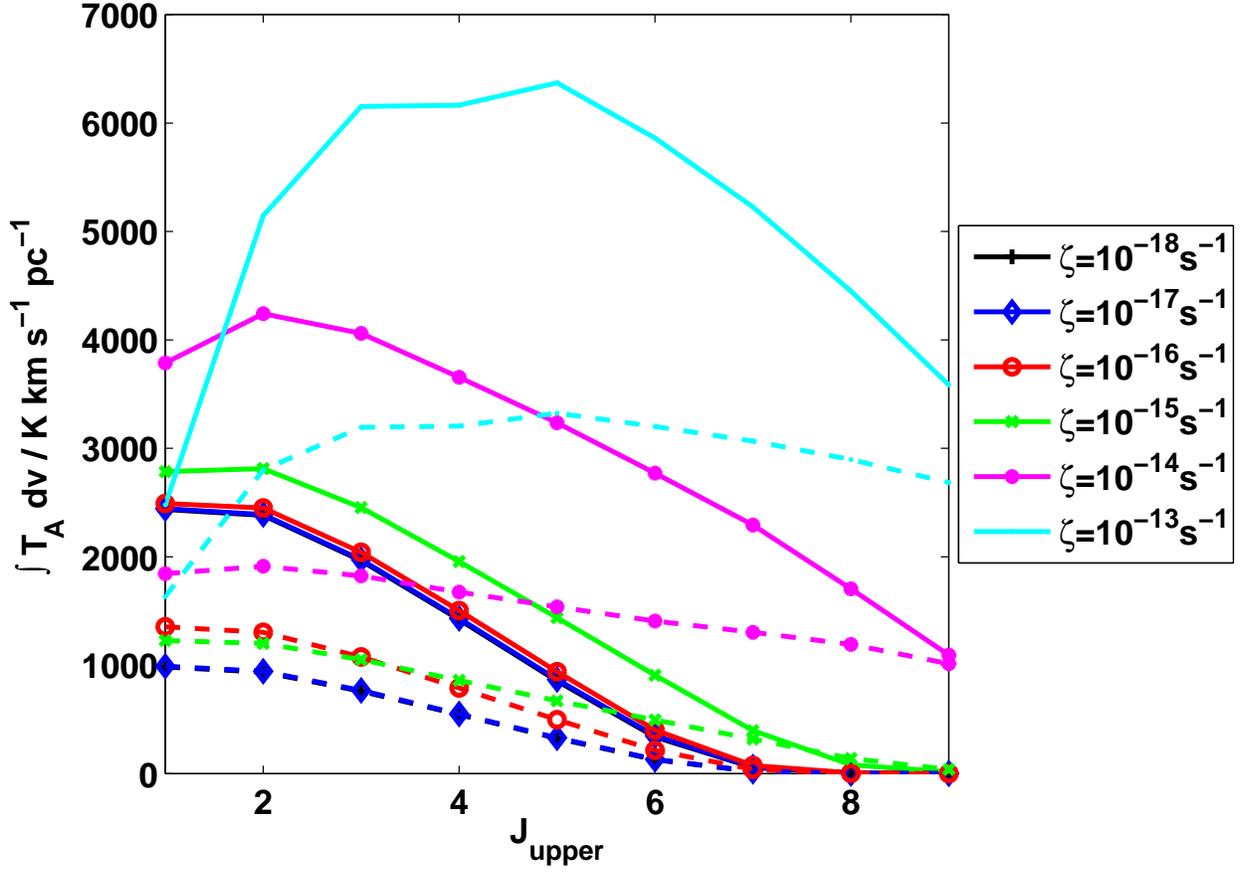}
\caption{Variation of relative velocity integrated CO antenna temperature per unit parsec with the cosmic ray ionisation rate at $A_v\sim$3 (solid lines) and $A_v \sim 8$ (dashed lines). The number density of hydrogen, metallicity and incident FUV radiation intensity are fixed at $n=10^5cm^{-3}$, $\xi=\xi_{0}$ and $\chi=1.7$.}
\label{fig:cr}
\end{figure}

The cosmic ray ionisation rate is varied between $\zeta$=10$^{-18}$ s$^{-1}$ and $\zeta$=10$^{-13}$ s$^{-1}$ with the rest of the input parameters fixed at $n$=10$^{5}$ cm$^{-3}$, $\xi=\xi_\odot$ and $\chi$=1.7 Draine. The results are shown in Figure \ref{fig:cr}.

 Apart from the highest cosmic-ray ionisation rates, the CO(1-0) line dominates the emission in almost all cases. The relative antenna temperature increases with increasing cosmic ray ionisation rate. At $\zeta=10^{-14}$ s$^{-1}$, the CO(2-1) transition just starts to dominate over the CO(1-0) line but by $\zeta=10^{-13}$ s$^{-1}$, the CO(5-4) line is now the brightest CO line. This line can therefore be used to trace high density systems at solar metallicity with extremely high cosmic-ray fluxes such as sources with AGN or a starburst nucleus. In B09 it was noted that systems such as this show some evidence for having top-heavy IMFs depending on the exact core temperature and magnetic field within these galaxies. In some active galaxies with physical conditions similar to that considered here, the magnetic pressure created through ambipolar diffusion may be sufficient to halt core collapse and formation of low-mass stars. 

\subsection{FUV Radiation Field Strength}

\label{sec:G}

\begin{figure}
\plotone{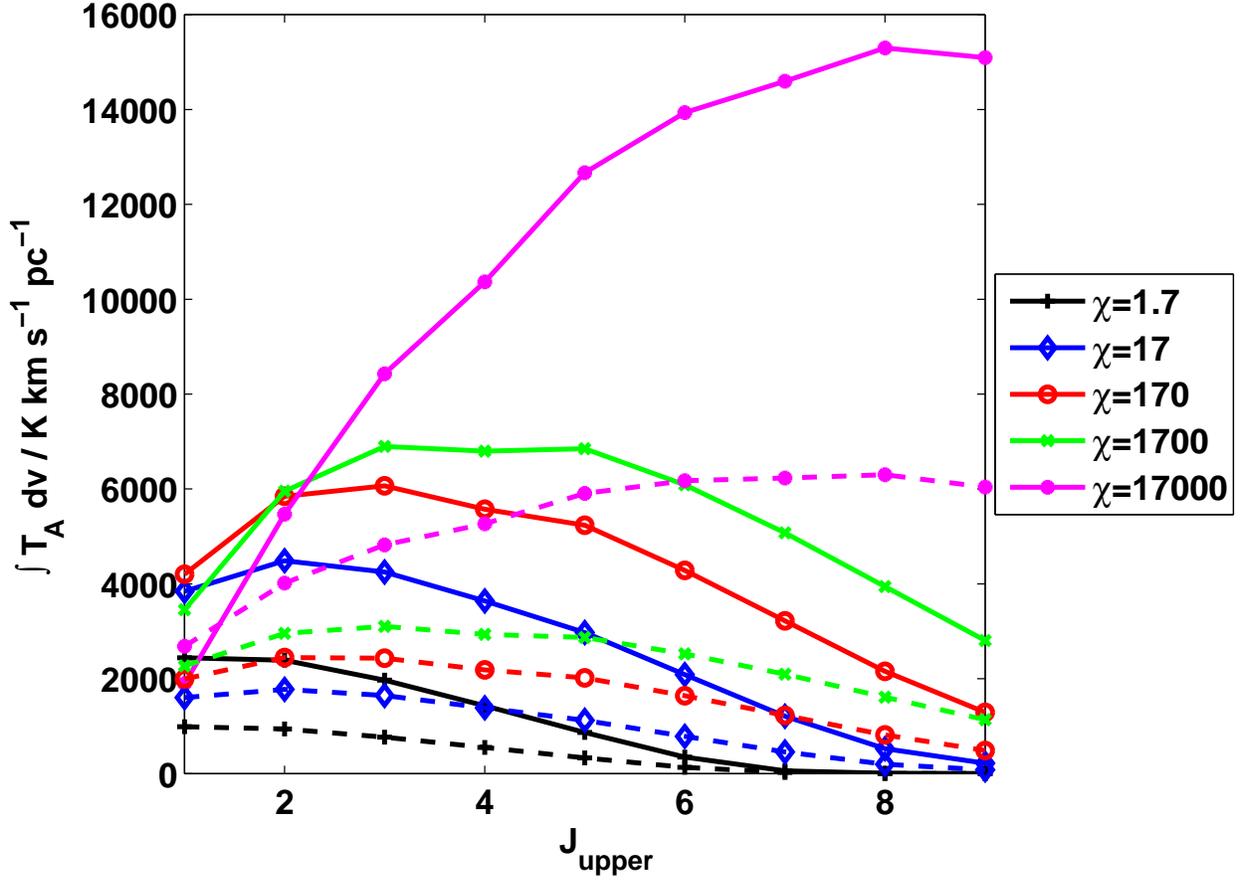}
\caption{Variation of relative velocity integrated CO antenna temperature per unit parsec with the FUV radiation field strength at $A_v\sim$3 (solid lines) and $A_v \sim 8$ (dashed lines). The number density of hydrogen, metallicity and cosmic ray ionisation rate are fixed at $n=10^5cm^{-3}$, $\xi=\xi_{0}$ and $\zeta=1\times10^{-17} s^{-1}$.}
\label{fig:G}
\end{figure}

The FUV radiation field strength is varied between $\chi$=1.7 Draine and $\chi$=1.7$\times10^4$ Draine with the rest of the input parameters fixed at $n$=10$^{5}$ cm$^{-3}$, $\xi=\xi_\odot$ and $\zeta$=10$^{-17}$ s$^{-1}$. The dependence of the CO SED on the FUV radiation field strength, is illustrated in Figure \ref{fig:G}. 

Increasing the FUV radiation field strength results in similar behaviour as increasing the cosmic ray ionisation rate as in B09. Higher-order lines become more and more dominant and the antenna temperature increases with $\chi$. At the maximum FUV field of $\xi=1.7 \times 10^4$ considered here, it is the CO(8-7) line that is dominant but for a FUV radiation field that is an order of magnitude weaker, the CO(3-2) line dominates. The high-order CO transitions can therefore be used as tracers of extremely high FUV radiation fields associated with the presence of massive stars. Such systems may also have top-heavy IMFs depending on whether the magnetic pressure is sufficient to halt the collapse and formation of low-mass stars. \\

Note that due to the large parameter space that we have tried to model, there will inevitably be degeneracies and the CO SEDs produced by models with $\chi=1700$ and $\zeta=10^{-14}$s$^{-1}$ may be observationally indistinguishable. However, we have associated large values for both these parameters with active galaxies producing massive stars in this rather simplistic analysis and it is probably fair to say that the form of the CO SED observed in our models can be taken to be roughly representative of such active systems.

\subsection{Trends in $^{12}$CO Line Emission at High-z}

\label{sec:imf}

We now consider the relative velocity integrated CO antenna temperatures for the high redshift models considered in B09. The physical parameters associated with these models are listed in Table \ref{tab:highz}. All models represent active galaxies with high cosmic ray fluxes and FUV radiation fields. Densities are also moderately high while both sub-solar and solar metallicities have been considered as the metallicity of high redshift systems still remains debated. B09 and references therein provide full justification for the choice of input parameters for these models. Once again we emphasise that these parameters are not meant to be representative of particular high redshift galaxies that have been observed in $^{12}$CO but rather crudely represent our best guess as to the parameters that are likely to be associated with high redshift systems. In B09 it was found that while there is no evidence for Models I or II having a high-mass biased IMF, Model III may have a high-mass biased IMF if the magnetic pressure is sufficient to halt the collapse and formation of low-mass stars. In other words, the physical conditions associated with Model III are likely to be necessary but not sufficient for a top-heavy IMF to exist. We consider differences in CO antenna temperatures in these three models as a possible means of distinguishing between the two IMF scenarios and the differing physical conditions in these galaxies. 

\begin{figure}
\begin{center}
\begin{minipage}[c]{1.00\textwidth} 
\centering 
\includegraphics[width=9cm,angle=0]{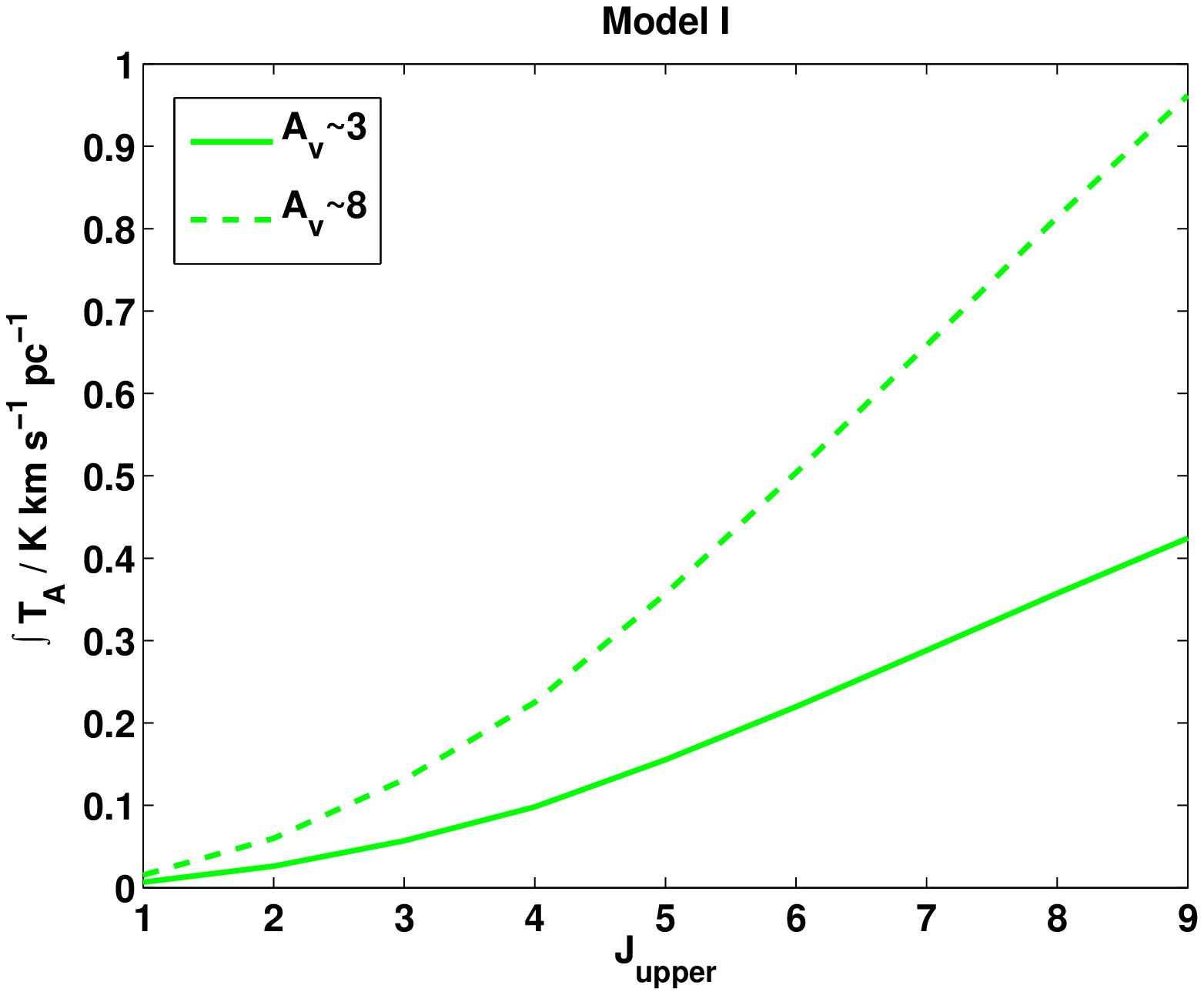}
\includegraphics[width=9cm,angle=0]{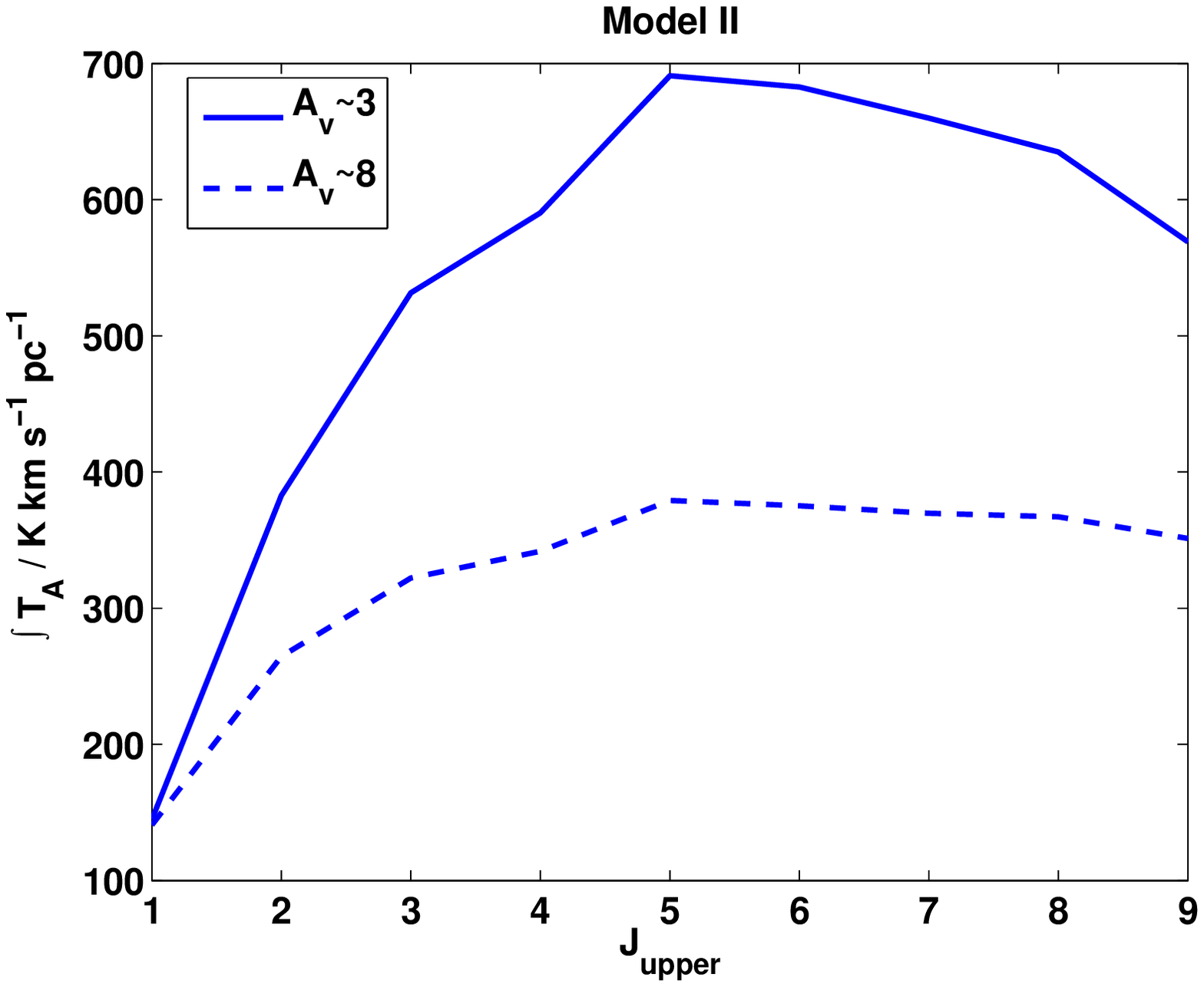}
\includegraphics[width=9cm,angle=0]{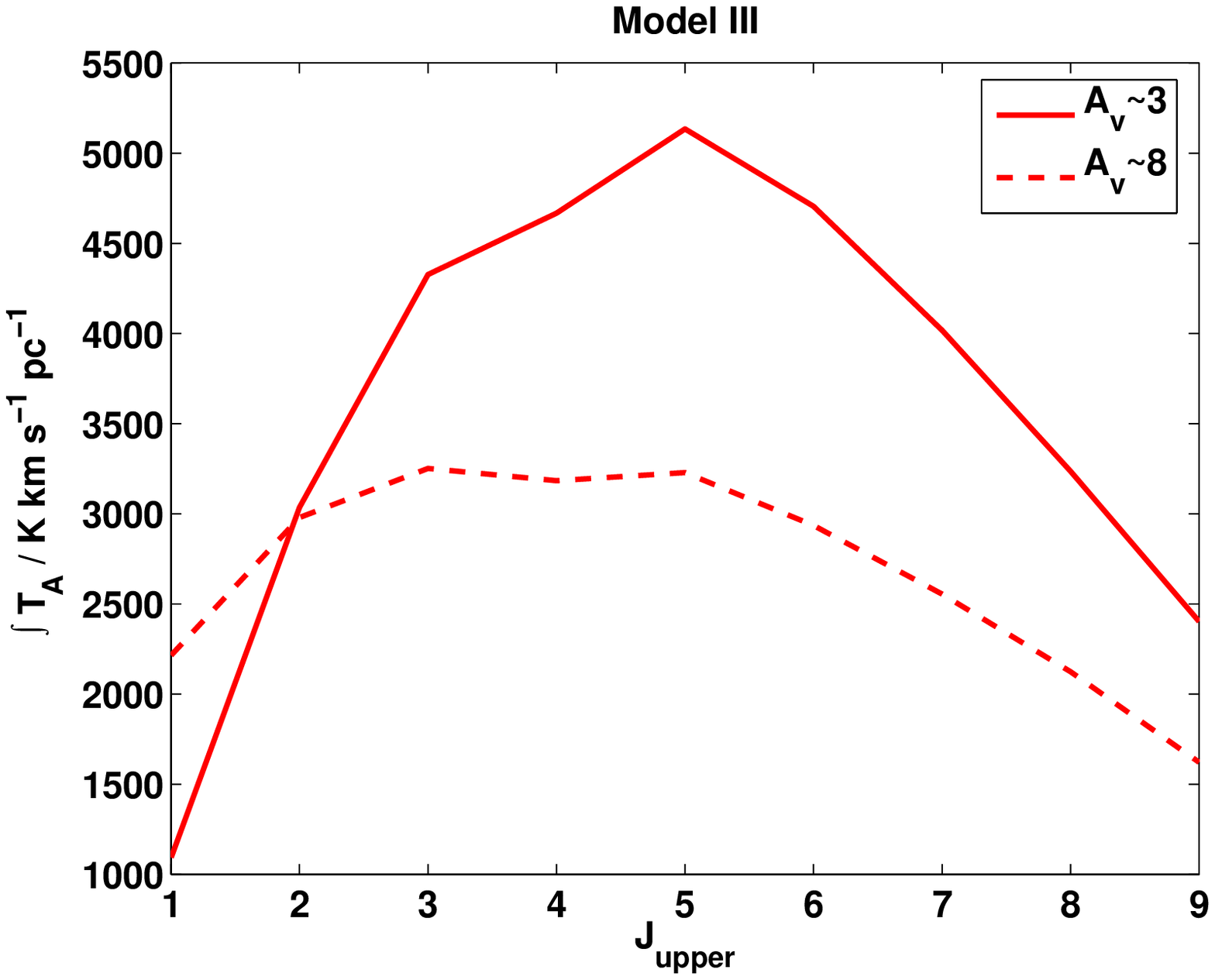}
\end{minipage}
\caption{Variation of theoretical velocity integrated CO antenna temperatures for high redshift models (Table \ref{tab:highz}) at $A_v\sim$3 (solid lines) and $A_v \sim 8$ (dashed lines).}
\label{fig:IMF1}
\end{center}
\end{figure}

Figure \ref{fig:IMF1} plots the relative velocity integrated CO antenna temperature for the different CO rotational lines for the three high-z models. Solid lines correspond to $A_v \sim 3$ while dashed lines correspond to $A_v \sim 8$. We can immediately see that the CO SEDs have very different forms for the three models.

Model I has very low velocity integrated antenna temperatures compared to the other two models. This is because the extreme cosmic ray fluxes and FUV fields invoked in this model destroy most of the CO, resulting in very low line intensities. The highest J transitions are found to be the brightest for this model but the temperatures associated with the lines are always very low probably indicating that such extreme conditions may be unphysical even in high redshift systems. Model II and Model III both have the same densities, cosmic ray fluxes and FUV fields, while Model II is at sub-solar metallicity, Model III has solar metallicity. The solar metallicity model is seen to produce brighter $\int (T_A)^{rel} dv$ than the sub-solar metallicity model. However, in both cases, it is the intermediate transitions that dominate the CO emission although the drop in the SED at high J is much sharper for the solar-metallicity model. At $A_v\sim$8, the CO SED for Model II peaks at J=5 before remaining roughly constant whereas the CO SED for Model III drops significantly at high J at the same $A_v$. These trends may be used to distinguish between solar and sub-solar metallicity active galaxies at high redshift. Figure \ref{fig:IMF1} clearly illustrates that the ratio of high-order CO transitions to the peak CO transition will be close to unity for physical conditions associated with an unbiased IMF at high redshift. However this ratio is likely to be smaller ($\sim0.5$) for physical conditions associated with a top-heavy IMF at high redshift. Unless such low CO J-ratios are seen in active galaxies at high redshift, an unbiased IMF cannot be ruled out. We caution the reader, however, that the physical conditions assumed in the models in this paper seem necessary but may not be sufficient to produce the IMFs associated with them in Table \ref{tab:highz}. Also, in $\S$ \ref{sec:xi}, we inferred that sub-solar metallicity counterparts of Milky Way type galaxies may also produce top-heavy IMFs but in this case the ratio of the high-J CO lines to the peak CO line is likely to be close to unity. Larger ratios of the CO high-J to peak transitions therefore seem to imply that the galaxy is at sub-solar metallicity and while such galaxies with normal levels of cosmic ray ionisation and FUV flux show evidence for producing top-heavy IMFs, the same galaxies at high-redshifts with elevated cosmic ray and FUV fluxes, may produce unbiased IMFs. The high-J to peak CO ratio therefore provides an useful indication of the metallicity of a galaxy and assuming it is known whether the galaxy is active, e.g., due to the presence of an AGN or starburst nucleus, the form of the IMF may be crudely inferred.

\begin{figure}
\plotone{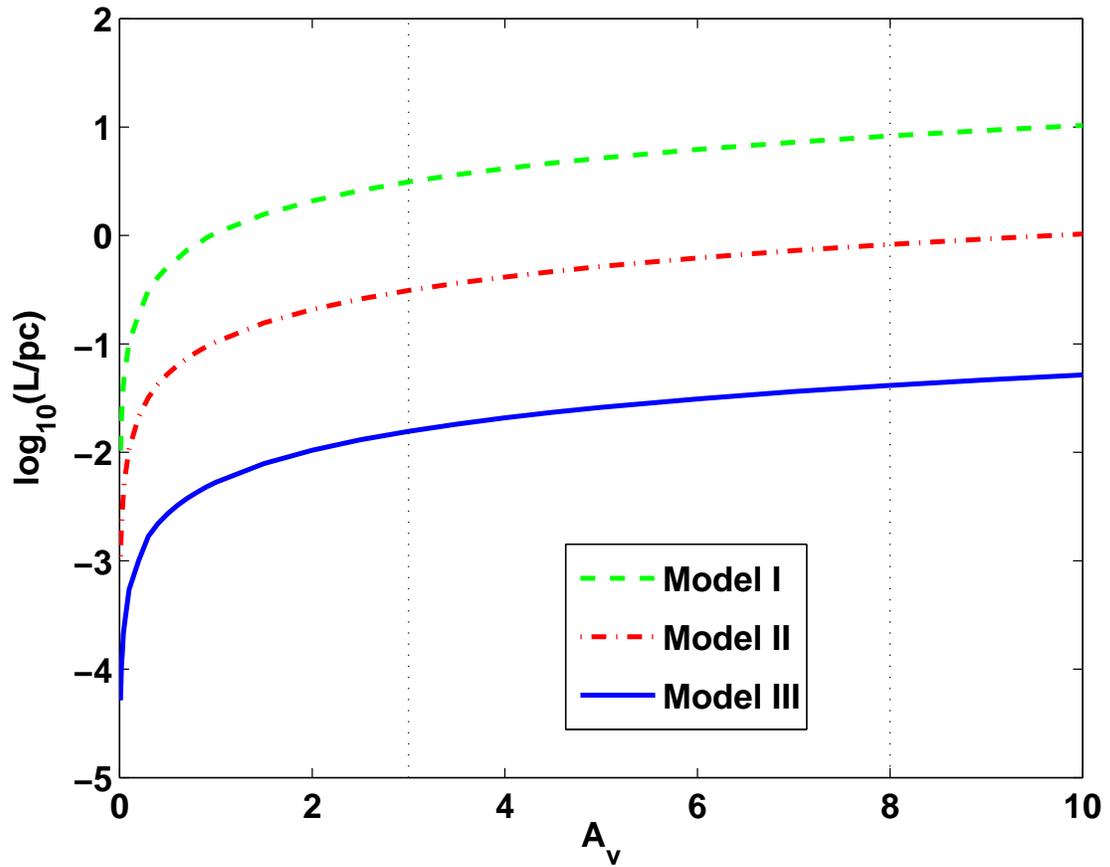}
\caption{Relationship between the size of the emitting region and the visual extinction, $A_v$ for our three high redshift models. The dotted lines correspond to the two $A_v$s at which we consider our outputs.}
\label{fig:size}
\end{figure}

In Figure \ref{fig:IMF2} we show variations in the theoretical velocity integrated antenna temperatures for the three models with increasing depth into the cloud. In all three models, the CO rotational lines get brighter with increasing depth within the cloud. Note that due to the differing number densities and metallicities of each model, the same $A_v$ corresponds to different physical distances for these models. Figure \ref{fig:size} illustrates the relationship between $A_v$ and size of the emitting region for our high redshift models.  

We note that the difference in the theoretical velocity integrated antenna temperatures of Model II (sub-solar metallicity and unbiased IMF) and Model III (solar metallicity, top-heavy IMF) is greater for the higher order CO transitions deep within the cloud. At $A_v \gtrsim 4$, the temperatures associated with the CO(1-0) line for Model II and Model III are almost the same. However, the emitting region in Model II is several orders of magnitude larger than in Model III. The CO(7-6) or CO(9-8) transition on the other hand produces markedly different antenna temperatures at all depths for the three high-redshift models.  

\subsection{Comparison to Observations}

\label{sec:obs}

As previously mentioned, $^{12}$CO line ratios are a good way of comparing our theoretical predictions directly to observations. In this section, we consider line ratios for two types of observed extragalactic sources using where available line intensities or integrated temperatures measured through the same beam size for calculating these ratios. Once again we emphasise that none of our models mimic the exact physical conditions in the galaxies described in this section. Rather, they may crudely represent the average conditions within a PDR in a galaxy of that type.    

\subsubsection{Nearby Starburst Galaxies - e.g. NGC253}

The nearby starburst galaxy NGC253 has been extensively mapped in $^{12}$CO emission by various authors \citep{Bayet:04,Bradford:n253cr,Gusten:06}. \citet{Bradford:n253cr} compute line intensities for various CO transitions corrected to a 15'' beam. We use these values from their Table 3 (Column 6) to derive line intensity ratios for a few of the CO lines and find observed line intensity ratios of $\frac{CO(2-1)}{CO(1-0)} \sim 8.0$ and $\frac{CO(7-6)}{CO(4-3)} \sim 3.3$. In B09, NGC253 was cited as an example of a galaxy with a high cosmic-ray ionisation rate due to the presence of a starburst nucleus. The cosmic ray ionisation rate for this galaxy is thought to be $\sim$ 800 times greater than in the Galaxy \citep{Bradford:n253cr}. The theoretical line intensity ratios derived from our model with a cosmic-ray ionisation rate of $\zeta = 10^{-14}$ s$^{-1}$ described in $\S$ \ref{sec:cr} are $\frac{CO(2-1)}{CO(1-0)} \sim 8.3$ and $\frac{CO(7-6)}{CO(4-3)} \sim 4.1$ at $A_v \sim 8$. Taking the ratio of intensities from Table 3 of \citet{Bayet:04} we find $\frac{CO(2-1)}{CO(1-0)} \sim 9.3$ for a beam size of 23'' \citep{Mauersberger:96, Harrison:99} in both cases and $\frac{CO(7-6)}{CO(4-3)} \sim 4.5$ for a beam size of 21.9'' \citep{Guesten:93} in both cases. The ratio of fluxes given in Table 1 of \citet{Gusten:06} gives $\frac{CO(7-6)}{CO(4-3)} \sim 2$. Given that we have not attempted to accurately model the physical conditions of this particular galaxy, the agreement between the observed and theoretical results is remarkably good. Note that these ratios are the ratios of the line intensities, $I$, rather than the velocity integrated temperature plotted in the figures as only the intensities in Table 3 of \citet{Bradford:n253cr} are all corrected to the same beam size. We note that the ratios of our velocity integrated temperatures agree with those in \citet{Bayet:04} to the same level as the intensity ratios. Our predicted line intensities and ratios also agree with the theoretical work of \citet{Meijerink:cr} (Tables 2, 3 and 4) who have looked at the effect of high cosmic-ray ionisation rates and FUV radiation fields on the $^{12}$CO lines

\subsubsection{High Redshift Galaxies - e.g. Cloverleaf}

The gravitationally lensed QSO, Cloverleaf (H1413+117) at a redshift of $\sim$ 2.5, was cited as an example of a high-redshift source in B09. \citet{Barvainis:97} report CO observations for this source and find the CO(4-3) line to be the strongest line in terms of the brightness temperature. These authors compute brightness temperature ratios of $\frac{CO(3-2)}{CO(4-3)} = 0.83 \pm 0.16$, $\frac{CO(5-4)}{CO(4-3)} = 0.73 \pm 0.16$ and $\frac{CO(7-6)}{CO(4-3)} = 0.68 \pm 0.13$ relative to the brightest line. There is therefore, a definite drop in the ratios relative to the brightest line when going from the low-J to the high-J transitions in this source. Figure \ref{fig:IMF1} shows that for Model II at both $A_v \sim 3$ and $A_v \sim 8$ as well as Model III at $A_v \sim 3$, the CO(5-4) line is the strongest line. For Model III at $A_v \sim 8$, the CO(3-2) line is marginally brighter than the CO(4-3) line and is now the strongest line. We compute ratios relative to the brightest line for both our high redshift models II and III. For Model II which shows little evidence for having a high-mass biased IMF, the ratios go from $\sim$0.7 for the low-J transitions to $\sim$1 for the CO(7-6) transition relative to the brighest line. For Model III which has some of the physical conditions necessary for producing a top-heavy IMF, the ratios go from $\sim$0.9 for the low-J transitions to $\sim$0.7 for the CO(7-6) line relative to the brightest line. Thus, the trends predicted by Model III seem to match the observations for this source better than those predicted by Model II. Once again, we emphasise that we have not tried, in our models, to match the physical conditions for this source in any way whatsoever or to try and reproduce the observed SED of this galaxy. However, our theoretical work does present some evidence, albeit speculative, that the Cloverleaf source at redshift $\sim$2.5 has roughly solar metallicity and potentially a top-heavy IMF.  It is expected that future observational studies will be able to test the validity of this prediction.


\section{Molecular Tracers of IMF at High-z}

\label{sec:abundances}

\begin{figure}[p]
\begin{center}
\begin{tabular}{cc}
\includegraphics[width=9cm,angle=0]{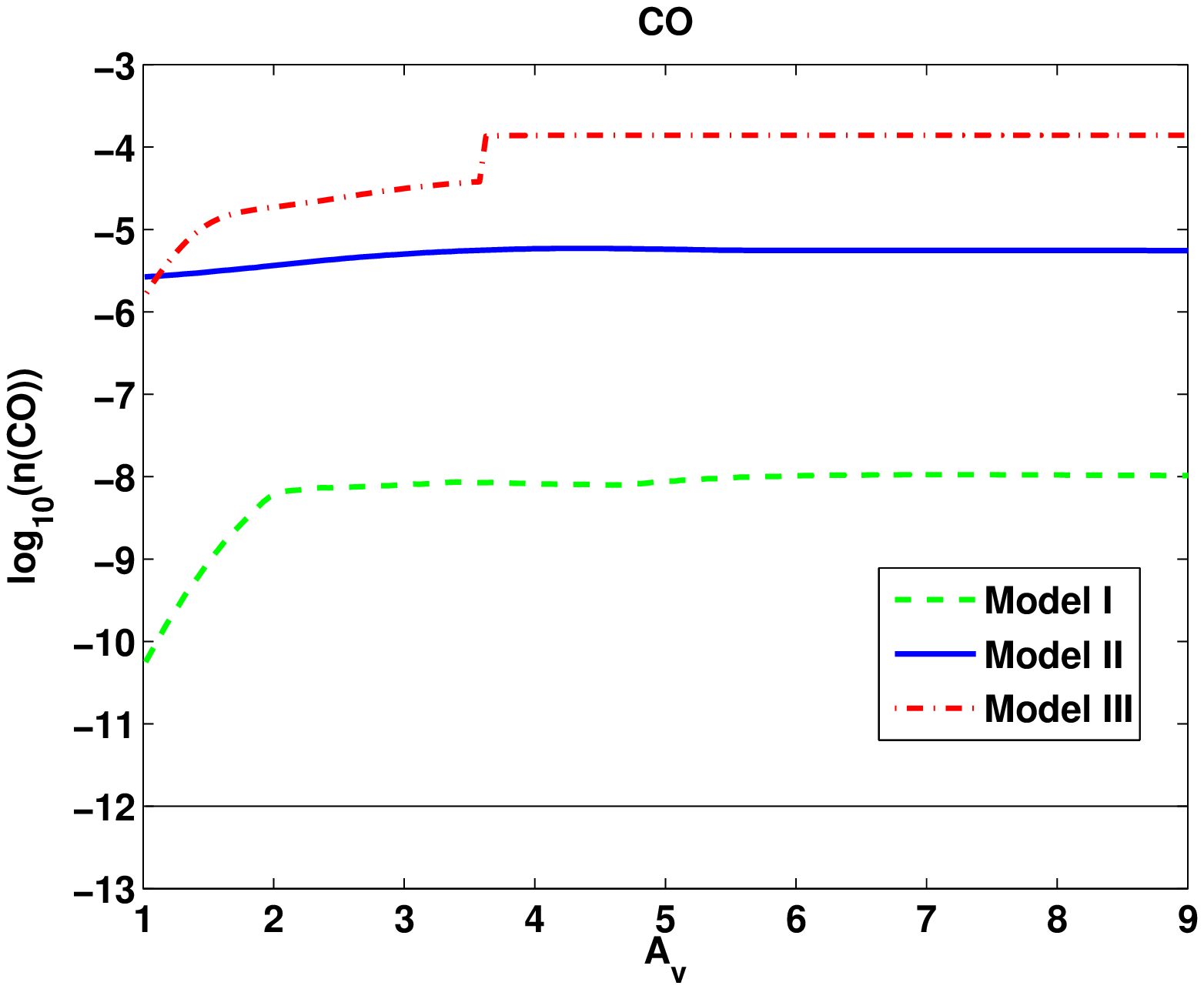} & \includegraphics[width=9cm,angle=0]{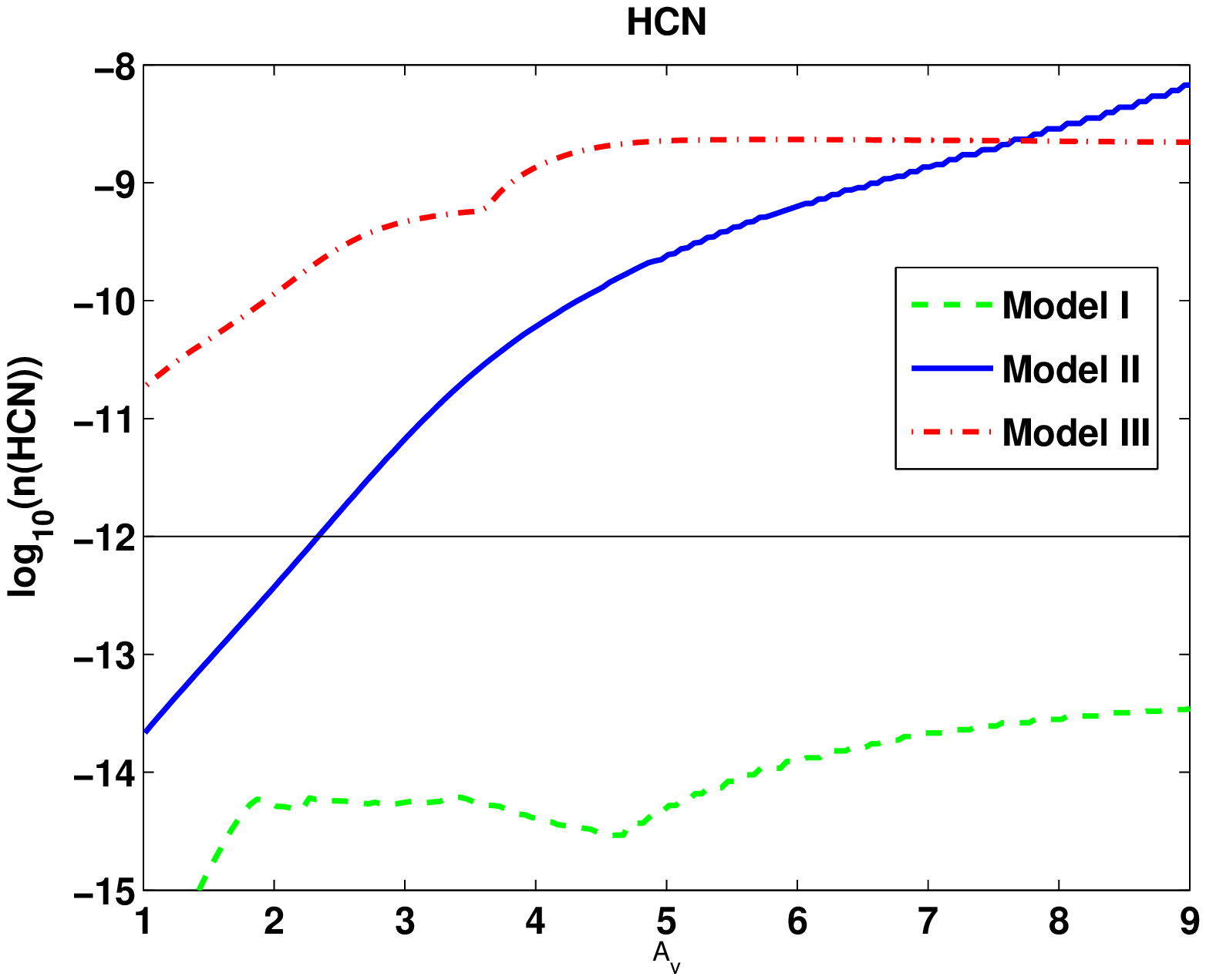} \\
\includegraphics[width=9cm,angle=0]{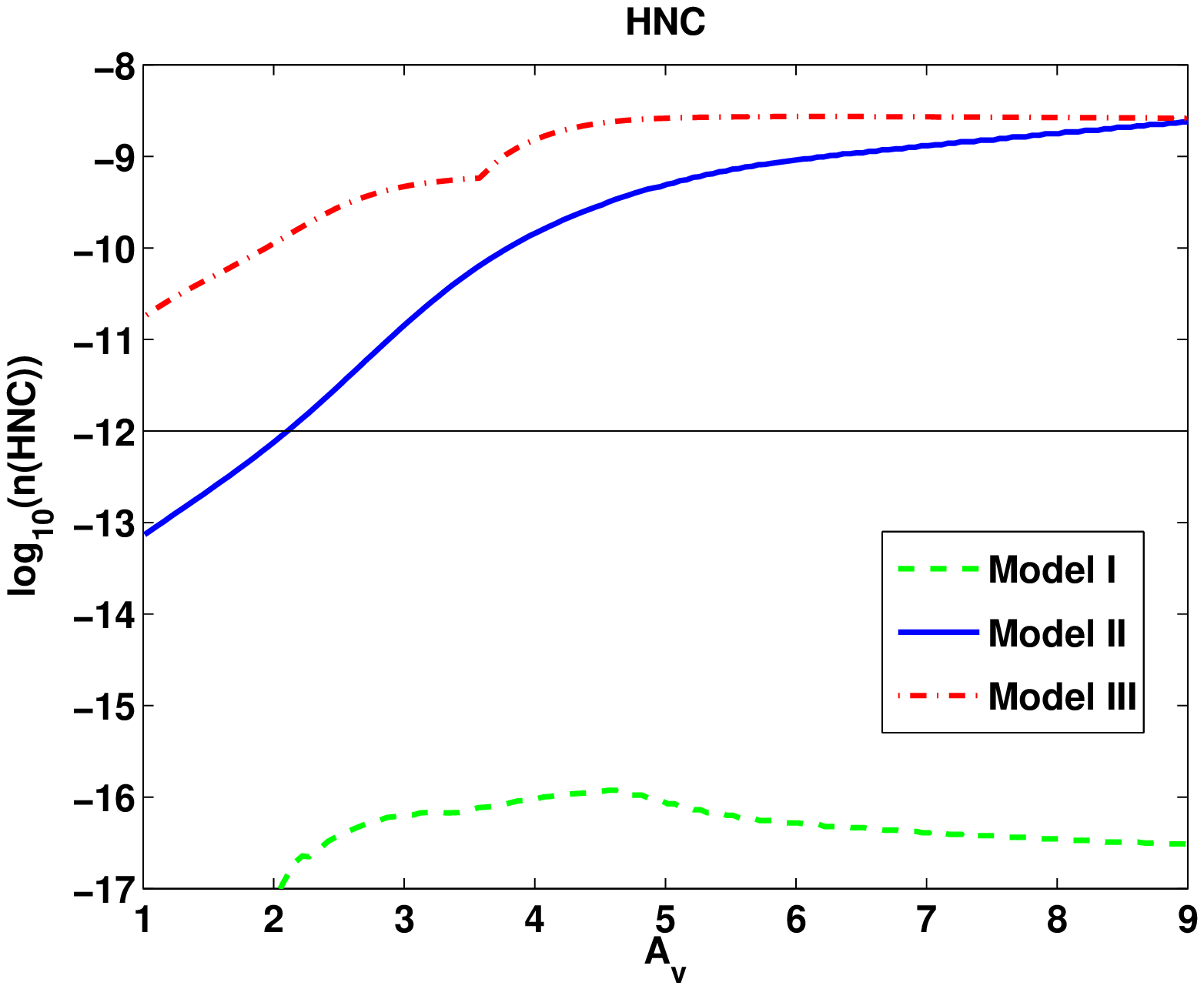} & \includegraphics[width=9cm,angle=0]{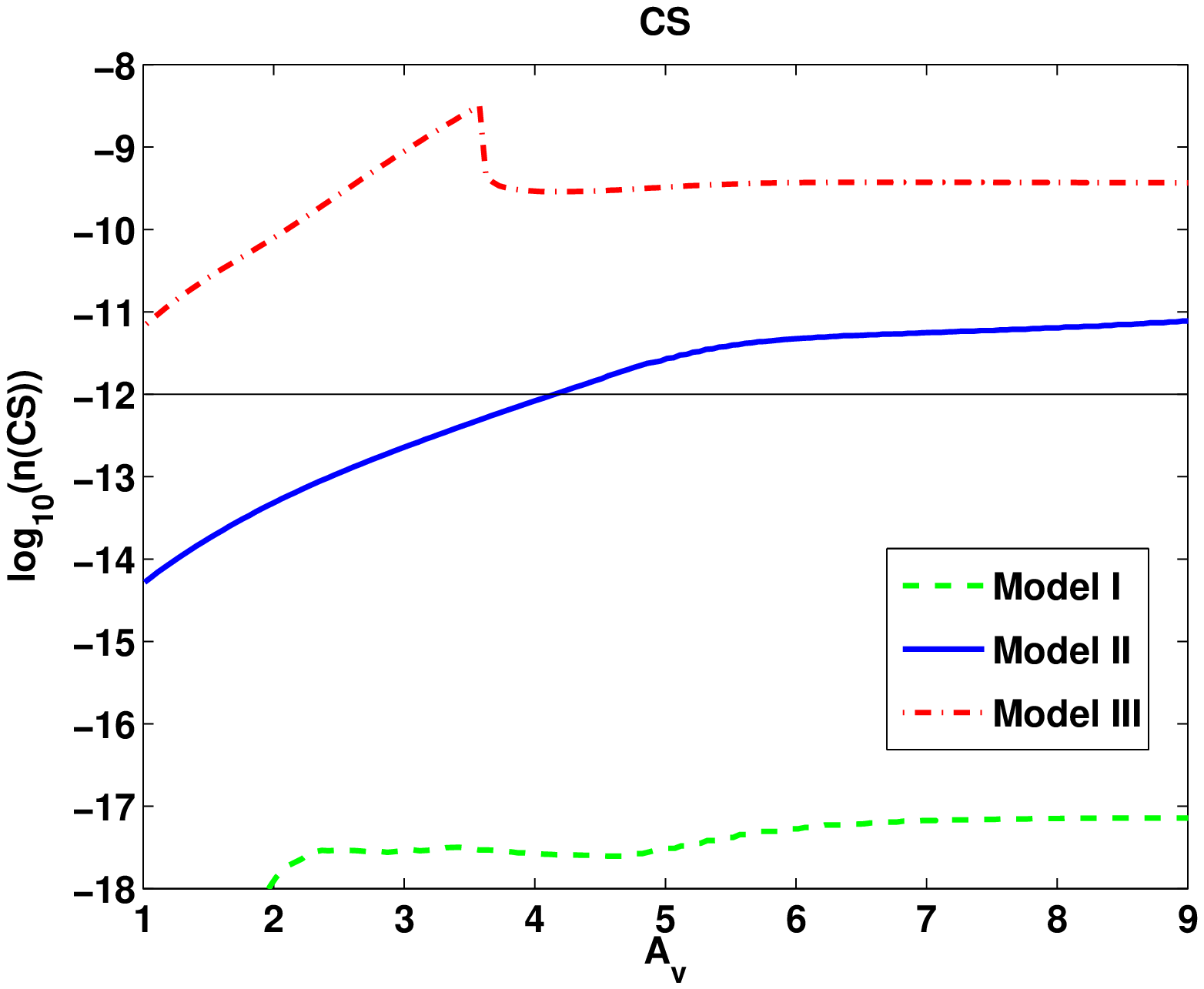} \\
\includegraphics[width=9cm,angle=0]{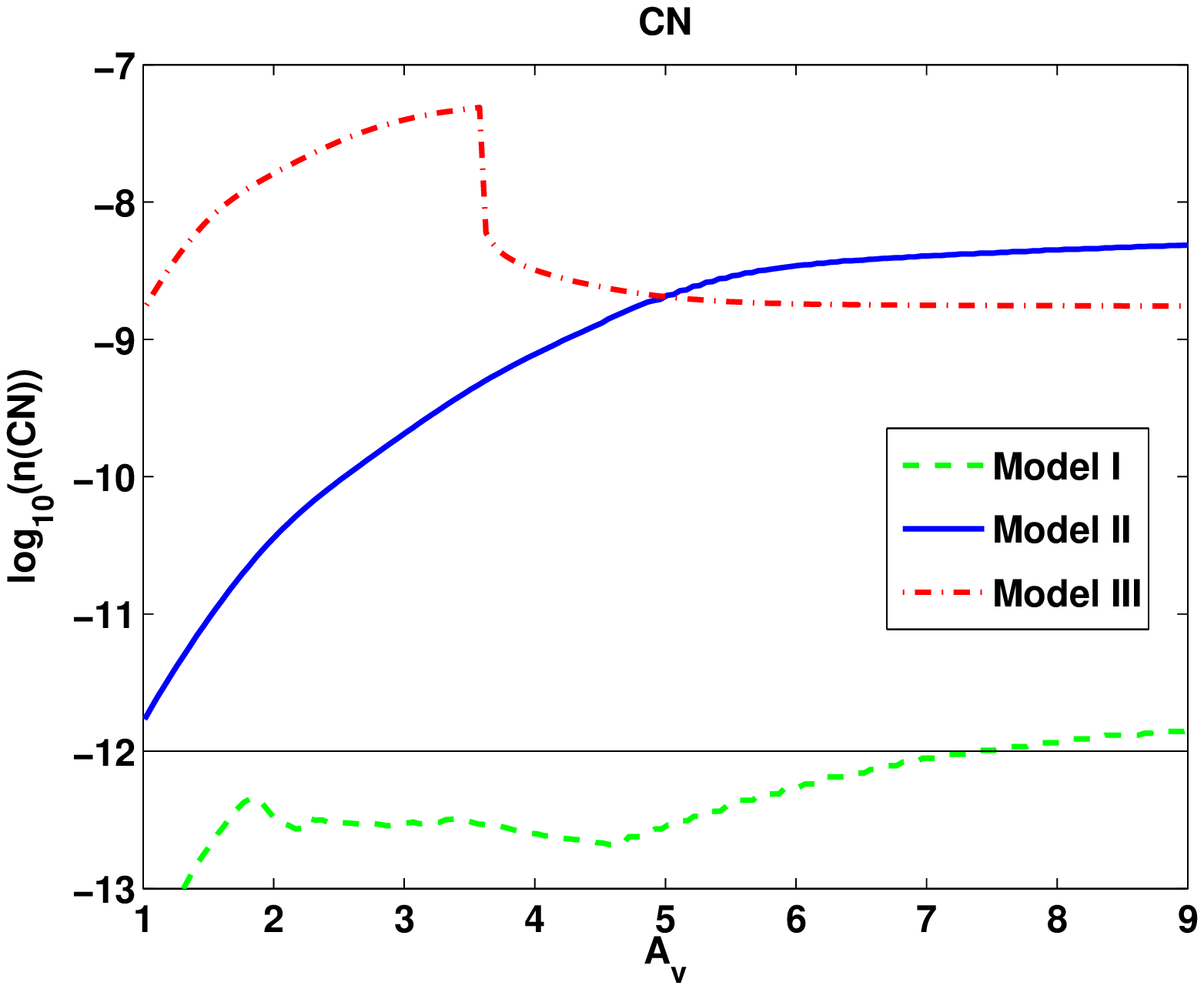} & \includegraphics[width=9cm,angle=0]{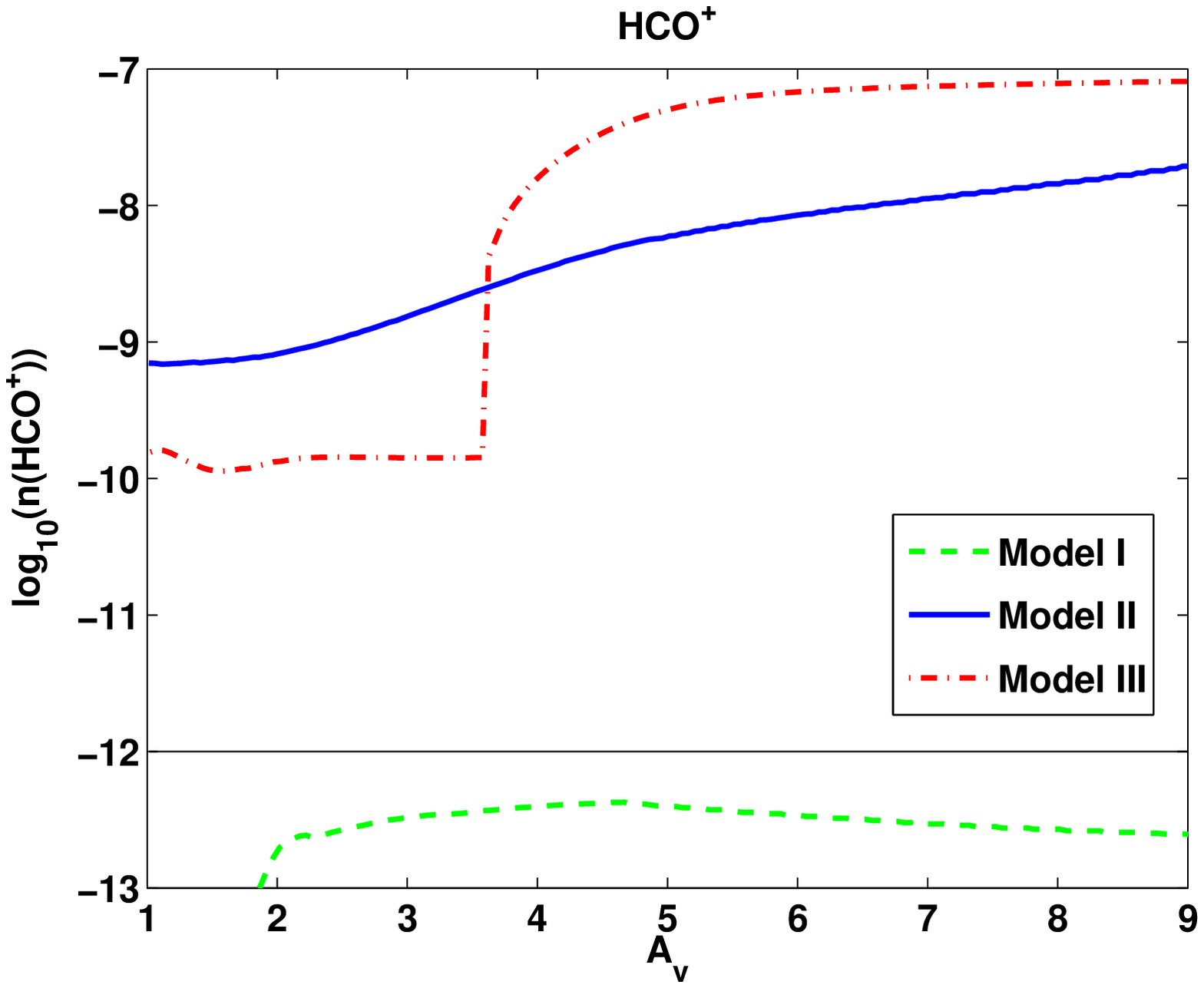} \\
\end{tabular}
\caption{Variation of fractional abundances of different molecular species with depth, $A_v$ for high-redshift models (Table \ref{tab:highz}). The solid black line indicates the assumed limit of detectability of 10$^{-12}$}
\label{fig:abundance}
\end{center}
\end{figure}

Although the focus of the paper so far has been on the $^{12}$CO SED associated with different physical conditions, the UCL\_PDR code includes a network of 83 chemical species whose abundances are also output by the code. In this section we consider how the fractional abundances of some of these different molecular species that could potentially be observed with future surveys, change for different IMFs at high redshift. The fractional abundance is defined to be the abundance of the molecule relative to the total hydrogen abundance, $n_H=n(H)+n(H_2)$. Once again we consider Models I, II and III of B09. In Table \ref{tab:mol} we list the fractional abundances of some key species for the three models both at $A_v \sim 3$ and $A_v \sim 8$ typical of the translucent gas component and dense PDR gas component of galaxies. In Figure \ref{fig:abundance} we also plot the variation of the molecular abundances of some key species with depth inside the cloud for the different high-redshift models.

From Table \ref{tab:mol} and Figure \ref{fig:abundance}, we observe the following trends. Assuming a limit of detectability of 10$^{-12}$ in fractional abundance which is arbitrary but roughly satisfied in our own galaxy \citep{Bayet:09}, all the species apart from CO are below this limit for Model I where the extremely high cosmic ray fluxes and FUV fields serve to destroy molecules effectively even deep within the cloud. Note that our qualitative results do not change if we assume a slightly higher or lower detectability limit. 

For Model III, the fractional abundances of all molecules considered here are above the detectability limit even at low $A_v$ where we normally do not expect complex molecules to be present. However, the high density in this model means that the systems considered here are always likely to be compact rather than diffuse and the size of the cloud is smaller than for any of the other high redshift models at a given $A_v$ as is illustrated by Figure \ref{fig:size}. The high fractional abundance of CO in this model also traces a high abundance of $H_2$ even at small $A_v$. 

While HCN and HNC are detectable in Model III at all $A_v$, these molecules only start becoming detectable in Model II at $A_v \sim 2$. However, the size of the emitting region in Model II is much larger than in Model III. At high $A_v$, the abundances of both HCN and HNC are almost the same in Models II and III corresponding to unbiased and high-mass biased IMFs respectively. 

The abundance of CO and CS is found to scale linearly with the metallicity - these abundances are about two orders of magnitude greater for Model III compared to Model II at all depths and the metallicity in Model III is also about two orders of magnitude greater than in Model II. A detection of CS at a much higher level than otherwise expected in a high redshift active galaxy, may therefore signal that it is at roughly solar metallicity and has some of the physical conditions necessary for producing a top-heavy IMF. However, we note that the predicted CS fractional abundance may be sensitive to the assumed initial abundance of sulphur in our models which is taken to be $2\times10^{-7}$ here. This fractional abundance reproduces roughly the observed abundance of CS seen in TMC-1 \citep{Ohishi:98} but is still highly uncertain \citep{Ruffle:99}. These results for the abundance of CO and CS are consistent with the findings of \citet{Bayet:09}. These authors have studied the molecular tracers of various PDR dominated galaxies with different physical conditions. For a density of 10$^4$ cm$^{-3}$, a cosmic ray ionisation rate of $5\times10^{-17}$ s$^{-1}$ and a FUV radiation field strength of $\chi=1.7\times10^3$, they find that CO and CS are linear tracers of the metallicity while CN and HCN are relatively insensitive to the metallicity. For $A_v \gtrsim 5$, the CN and HCN abundance are indeed found to be roughly the same for Models II and III. However, at lower $A_v$, both CN and HCN are much more abundant in Model III. 

The trend of HCO$^+$ with $A_v$ provides another possibly interesting diagnostic for different IMFs at high redshift. This molecule is more abundant at low $A_v$ ($A_v \lesssim 3.5$) in the sub-solar metallicity model which is thought to produce an unbiased IMF. However at $A_v \gtrsim 3.5$ it becomes more abundant in the solar metallicity model which is thought to produce a top-heavy IMF. For Model III, there is an obvious change in the chemistry at $A_v \sim 3.5$ and a sharp change in the abundance of all the molecules is seen at this $A_v$. It is well known that the attenuation of the FUV flux with distance in a photon-dominated region, produces a characteristic depth dependent chemical structure with well-defined chemical zones. At $A_v \sim 3.5$ in this model, we witness the transition from a cloud layer of singly ionized carbon produced by photo-ionization of carbon atoms by the FUV photons, to a layer where most of the atomic carbon becomes locked up in the stable molecule CO due to the inability of the FUV field to penetrate the cloud further. It should be noted that in our other high redshift models, which are both at sub-solar metallicities, this transition between chemical zones is not seen to occur within the $A_v$ range considered in Figure \ref{fig:abundance}.

Although the variation of molecular fractional abundances with $A_v$ demonstrates some interesting differences between the various IMF models, in reality these trends may be hard to observe. Most observations of extragalactic sources lead to measurements of the total column density of different molecules integrated over the depth of the cloud. Therefore, the observed column density of molecules such as HCO$^+$ for example in two extragalactic sources with different IMFs may actually be very similar and therefore not a good diagnostic for the form of the IMF.

\section{Discussion \& Conclusions}

\label{sec:disc}

If top-heavy IMFs arise from the suppression of low-mass star formation
then the galactic physical conditions that inhibit star formation may be 
crudely defined. Physical conditions in which low-mass star formation is 
permitted are here assumed to give rise to unbiased IMFs. Since the 
physical conditions determine the chemistry, potential molecular 
diagnostics of various types of IMF can be deduced.

The sensitivity of the $^{12}$CO rotational spectrum to changes in gas number 
density, metallicity, and cosmic ray and far-UV fluxes are presented and 
used as a basis for predicting the relative velocity-integrated antenna 
temperature of the CO rotational spectrum in three models of high redshift 
active galaxies defined in B09. While active galaxies with near-solar 
metallicity (and top-heavy IMFs) produce higher antenna temperatures than 
those with sub-solar metallicity (and unbiased IMFs) the actual 
rotational distribution is fairly similar. In sub-solar metallicity counterparts of Milky Way type galaxies, the high-J to peak CO line ratio approaches unity. These physical conditions are thought to produce top-heavy IMFs. However, sub-solar metallicity active galaxies at high redshift also have high-J to peak CO line ratios that approach unity and these physical conditions are thought to produce unbiased IMFs. Solar metallicity active galaxies at high redshift have high-J to peak CO ratios that drop well below unity and these conditions may produce top-heavy IMFs. The form of the CO SED therefore clearly depends strongly on the metallicity of the galaxy and provided it is known whether the galaxy is active or quiescent, the form of the IMF that may be produced under such physical conditions, can be crudely inferred. High order CO transitions are also found to 
provide a good diagnostic for high far-UV intensity counterparts of Milky Way type systems which show some evidence for producing high-mass biased stellar IMFs. 

It is shown that the theoretical predictions for the $^{12}$CO rotational spectrum can be matched to observed line ratios of different kinds of extragalactic sources both nearby and in the distant Universe. Potentially exciting new evidence is presented for the Cloverleaf QSO at z$\sim2.5$ having the physical conditions necessary for producing a top-heavy IMF. This is consistent with the studies of \citet{Dave:08, Wilkins:IMF} and \citet{VanDokkum:08} who require the IMF to be top-heavy at $z>1$.

Molecular tracers of dense gas are predicted to differ markedly between 
galaxies with top-heavy and unbiased IMFs. In particular, CO and CS 
differ significantly over the $A_v$ range 1 to 9 visual magnitudes with these molecules acting as 
linear tracers of the metallicity. As solar metallicity active galaxies at high redshift are thought to have physical conditions necessary for a top-heavy IMF, detection of elevated levels of CO and CS in high-redshift active galaxies may signal that such physical conditions exist in these sources. HCN, HNC and CN differ between solar and sub-solar metallicity active galaxies at high redshift 
for small visual extinctions but at large depths, these molecules are found to be independent of the IMF shape. Finally, HCO$^+$ is more
abundant in active galaxies with unbiased IMFs at low $A_v$ but, at high $A_v$, the abundance of this molecule is greater
in active galaxies with  top-heavy IMFs. Unless the $A_v$ of the host galaxy can be directly measured, this molecule is probably not a good diagnostic for the shape of the IMF as the integrated HCO$^+$ column density that would arise from our two high-redshift models with different IMFs, would probably be very similar. 

We conclude that chemistry may in some cases provide a simple and independent means of 
determining the approximate nature of the IMF of a high redshift galaxy.

\section*{Acknowledgements}

We thank the referee for a useful and constructive report that has helped improve this paper. We would also like to thank Estelle Bayet for useful discussions and Jeremy Yates for his help with cluster computing. MB is supported by an STFC studentship. 

\newpage
\begin{landscape}
\begin{table}
\caption{High Redshift Models: inputs to the model as well as the form of the IMF associated with these physical conditions and inferred in B09. Note that these physical conditions seem necessary but may not be sufficient for producing the IMFs associated with them.} \label{tab:highz}
  \begin{center}
    \begin{tabular}{c|c|c|c|c|c}
      \bf{Model} & \bf{Density (cm$^{-3}$)} & \bf{Metallicity ($\xi_\odot$)} & \bf{CR Ionisation Rate (s$^{-1}$)} & \bf{FUV Flux (Draine)} & \bf{IMF}\\
      \hline
      I & 10$^4$ & 0.05 & 10$^{-13}$ & 1.7$\times10^{4}$ & Unbiased \\
      II & 10$^5$ & 0.05 & 10$^{-14}$ & 1.7$\times10^{3}$ & Unbiased \\
      III & 10$^5$ & 1.00 & 10$^{-14}$ & 1.7$\times10^{3}$ & Top-heavy \\
      \hline
    \end{tabular}    \vspace{2mm}
  \end{center}
\end{table}
\end{landscape}


\begin{landscape}
\begin{table}
  \caption{Fractional Molecular Abundances for different IMF Models} \label{tab:mol}
  \begin{center}
    \begin{tabular}{c| cc | cc | cc}
      \textbf{Molecule} & \multicolumn{2}{c|}{\textbf{Model I}} & \multicolumn{2}{|c}{\textbf{Model II}} & \multicolumn{2}{|c}{\textbf{Model III}} \\
      \hline
       & $A_v\sim3$ & $A_v\sim8$ & $A_v\sim3$ & $A_v\sim8$  & $A_v\sim3$ & $A_v\sim8$  \\
      \hline \hline
      HCN & $5.598\times10^{-15}$ & $2.813\times10^{-14}$ & $6.931\times10^{-12}$ & $2.869\times10^{-9}$ & $4.753\times10^{-10}$ & $2.253\times10^{-9}$ \\
      HNC & $6.284\times10^{-17}$ & $3.498\times10^{-17}$ & $1.475\times10^{-11}$ & $1.770\times10^{-9}$ & $4.753\times10^{-10}$ & $2.655\times10^{-9}$ \\
      CS & $2.905\times10^{-18}$ & $7.077\times10^{-18}$ & $2.316\times10^{-13}$ & $6.405\times10^{-12}$ & $9.496\times10^{-10}$ & $3.719\times10^{-10}$ \\
      CN & $3.008\times10^{-13}$ & $1.153\times10^{-12}$ & $2.094\times10^{-10}$ & $4.474\times10^{-9}$ & $4.029\times10^{-8}$ & $1.758\times10^{-9}$ \\
      HCO$^+$ & $3.279\times10^{-13}$ & $2.685\times10^{-13}$ & $1.553\times10^{-9}$ & $1.438\times10^{-8}$ & $1.413\times10^{-10}$ & $7.832\times10^{-8}$ \\
      CO & $8.037\times10^{-9}$ & $1.050\times10^{-8}$ & $5.063\times10^{-6}$ & $5.591\times10^{-6}$ & $3.183\times10^{-5}$ & $1.397\times10^{-4}$ \\
      \hline
    \end{tabular}    \vspace{2mm}
  \end{center}
\end{table}
\end{landscape}

\newpage

\bibliography{}

\begin{thebibliography}{32}
\expandafter\ifx\csname natexlab\endcsname\relax\def\natexlab#1{#1}\fi

\bibitem[{{Banerji} {et~al.}(2009){Banerji}, {Viti}, {Williams}, \&
  {Rawlings}}]{Banerji:timescales}
{Banerji}, M., {Viti}, S., {Williams}, D.~A., \& {Rawlings}, J.~M.~C. 2009,
  \apj, 692, 283

\bibitem[{{Barvainis} {et~al.}(1997){Barvainis}, {Maloney}, {Antonucci}, \&
  {Alloin}}]{Barvainis:97}
{Barvainis}, R., {Maloney}, P., {Antonucci}, R., \& {Alloin}, D. 1997, \apj,
  484, 695

\bibitem[{{Bayet} {et~al.}(2004){Bayet}, {Gerin}, {Phillips}, \&
  {Contursi}}]{Bayet:04}
{Bayet}, E., {Gerin}, M., {Phillips}, T.~G., \& {Contursi}, A. 2004, \aap, 427,
  45

\bibitem[{{Bayet} {et~al.}(2006){Bayet}, {Gerin}, {Phillips}, \&
  {Contursi}}]{Bayet:06}
---. 2006, \aap, 460, 467

\bibitem[{{Bayet} {et~al.}(2009){Bayet}, {Viti}, {Williams}, {Rawlings}, \&
  {Bell}}]{Bayet:09}
{Bayet}, E., {Viti}, S., {Williams}, D.~A., {Rawlings}, J.~M.~C., \& {Bell}, T.
  2009, \apj, 696, 1466

\bibitem[{{Bell} {et~al.}(2006){Bell}, {Roueff}, {Viti}, \&
  {Williams}}]{Bell:Xfactor1}
{Bell}, T.~A., {Roueff}, E., {Viti}, S., \& {Williams}, D.~A. 2006, \mnras,
  371, 1865

\bibitem[{{Bell} {et~al.}(2007){Bell}, {Viti}, \& {Williams}}]{Bell:Xfactor2}
{Bell}, T.~A., {Viti}, S., \& {Williams}, D.~A. 2007, \mnras, 378, 983

\bibitem[{{Bertoldi} {et~al.}(2003){Bertoldi}, {Cox}, {Neri}, {Carilli},
  {Walter}, {Omont}, {Beelen}, {Henkel}, {Fan}, {Strauss}, \&
  {Menten}}]{Bertoldi:03}
{Bertoldi}, F., {Cox}, P., {Neri}, R., {Carilli}, C.~L., {Walter}, F., {Omont},
  A., {Beelen}, A., {Henkel}, C., {Fan}, X., {Strauss}, M.~A., \& {Menten},
  K.~M. 2003, \aap, 409, L47

\bibitem[{{Bradford} {et~al.}(2003){Bradford}, {Nikola}, {Stacey}, {Bolatto},
  {Jackson}, {Savage}, {Davidson}, \& {Higdon}}]{Bradford:n253cr}
{Bradford}, C.~M., {Nikola}, T., {Stacey}, G.~J., {Bolatto}, A.~D., {Jackson},
  J.~M., {Savage}, M.~L., {Davidson}, J.~A., \& {Higdon}, S.~J. 2003, \apj,
  586, 891

\bibitem[{{Daddi} {et~al.}(2009){Daddi}, {Dannerbauer}, {Krips}, {Walter},
  {Dickinson}, {Elbaz}, \& {Morrison}}]{Daddi:COhighz}
{Daddi}, E., {Dannerbauer}, H., {Krips}, M., {Walter}, F., {Dickinson}, M.,
  {Elbaz}, D., \& {Morrison}, G.~E. 2009, ArXiv e-prints

\bibitem[{{Dame} {et~al.}(2001){Dame}, {Hartmann}, \& {Thaddeus}}]{Dame:01}
{Dame}, T.~M., {Hartmann}, D., \& {Thaddeus}, P. 2001, \apj, 547, 792

\bibitem[{{Dav{\'e}}(2008)}]{Dave:08}
{Dav{\'e}}, R. 2008, \mnras, 385, 147

\bibitem[{{de Jong} {et~al.}(1980){de Jong}, {Boland}, \&
  {Dalgarno}}]{deJong:80}
{de Jong}, T., {Boland}, W., \& {Dalgarno}, A. 1980, \aap, 91, 68

\bibitem[{{Draine}(1978)}]{Draine:ISRF}
{Draine}, B.~T. 1978, \apjs, 36, 595

\bibitem[{{Flower} \& {Launay}(1985)}]{Flower:85}
{Flower}, D.~R. \& {Launay}, J.~M. 1985, \mnras, 214, 271

\bibitem[{{Green} \& {Chapman}(1978)}]{Green:78}
{Green}, S. \& {Chapman}, S. 1978, \apjs, 37, 169

\bibitem[{{Green} \& {Thaddeus}(1976)}]{Green:76}
{Green}, S. \& {Thaddeus}, P. 1976, \apj, 205, 766

\bibitem[{{Guesten} {et~al.}(1993){Guesten}, {Serabyn}, {Kasemann},
  {Schinckel}, {Schneider}, {Schulz}, \& {Young}}]{Guesten:93}
{Guesten}, R., {Serabyn}, E., {Kasemann}, C., {Schinckel}, A., {Schneider}, G.,
  {Schulz}, A., \& {Young}, K. 1993, \apj, 402, 537

\bibitem[{{G{\"u}sten} {et~al.}(2006){G{\"u}sten}, {Philipp}, {Wei{\ss}}, \&
  {Klein}}]{Gusten:06}
{G{\"u}sten}, R., {Philipp}, S.~D., {Wei{\ss}}, A., \& {Klein}, B. 2006, \aap,
  454, L115

\bibitem[{{Harrison} {et~al.}(1999){Harrison}, {Henkel}, \&
  {Russell}}]{Harrison:99}
{Harrison}, A., {Henkel}, C., \& {Russell}, A. 1999, \mnras, 303, 157

\bibitem[{{Knudsen} {et~al.}(2009){Knudsen}, {Neri{\'e}}, {Kneib}, \& {van der
  Werf}}]{Knudsen:COhighz}
{Knudsen}, K.~K., {Neri{\'e}}, R., {Kneib}, J.-P., \& {van der Werf}, P.~P.
  2009, \aap, 496, 45

\bibitem[{{Le Teuff} {et~al.}(2000){Le Teuff}, {Millar}, \&
  {Markwick}}]{LeTeuff:UMIST99}
{Le Teuff}, Y.~H., {Millar}, T.~J., \& {Markwick}, A.~J. 2000, \aaps, 146, 157

\bibitem[{{Mauersberger} {et~al.}(1996){Mauersberger}, {Henkel}, {Wielebinski},
  {Wiklind}, \& {Reuter}}]{Mauersberger:96}
{Mauersberger}, R., {Henkel}, C., {Wielebinski}, R., {Wiklind}, T., \&
  {Reuter}, H.-P. 1996, \aap, 305, 421

\bibitem[{{Meijerink} {et~al.}(2006){Meijerink}, {Spaans}, \&
  {Israel}}]{Meijerink:cr}
{Meijerink}, R., {Spaans}, M., \& {Israel}, F.~P. 2006, \apjl, 650, L103

\bibitem[{{Nesvadba} {et~al.}(2009){Nesvadba}, {Neri}, {De Breuck}, {Lehnert},
  {Downes}, {Walter}, {Omont}, {Boulanger}, \& {Seymour}}]{Nesvadba:COhighz}
{Nesvadba}, N.~P.~H., {Neri}, R., {De Breuck}, C., {Lehnert}, M.~D., {Downes},
  D., {Walter}, F., {Omont}, A., {Boulanger}, F., \& {Seymour}, N. 2009,
  \mnras, L208+

\bibitem[{{Ohishi} \& {Kaifu}(1998)}]{Ohishi:98}
{Ohishi}, M. \& {Kaifu}, N. 1998, in Chemistry and Physics of Molecules and
  Grains in Space. Faraday Discussions No. 109, 205--+

\bibitem[{{R{\"o}llig} {et~al.}(2007){R{\"o}llig}, {Abel}, {Bell}, {Bensch},
  {Black}, {Ferland}, {Jonkheid}, {Kamp}, {Kaufman}, {Le Bourlot}, {Le Petit},
  {Meijerink}, {Morata}, {Ossenkopf}, {Roueff}, {Shaw}, {Spaans}, {Sternberg},
  {Stutzki}, {Thi}, {van Dishoeck}, {van Hoof}, {Viti}, \&
  {Wolfire}}]{Rollig:PDR}
{R{\"o}llig}, M., {Abel}, N.~P., {Bell}, T., {Bensch}, F., {Black}, J.,
  {Ferland}, G.~J., {Jonkheid}, B., {Kamp}, I., {Kaufman}, M.~J., {Le Bourlot},
  J., {Le Petit}, F., {Meijerink}, R., {Morata}, O., {Ossenkopf}, V., {Roueff},
  E., {Shaw}, G., {Spaans}, M., {Sternberg}, A., {Stutzki}, J., {Thi}, W.-F.,
  {van Dishoeck}, E.~F., {van Hoof}, P.~A.~M., {Viti}, S., \& {Wolfire}, M.~G.
  2007, \aap, 467, 187

\bibitem[{{Ruffle} {et~al.}(1999){Ruffle}, {Hartquist}, {Caselli}, \&
  {Williams}}]{Ruffle:99}
{Ruffle}, D.~P., {Hartquist}, T.~W., {Caselli}, P., \& {Williams}, D.~A. 1999,
  \mnras, 306, 691

\bibitem[{{Sakamoto}(1999)}]{Sakamoto:99}
{Sakamoto}, S. 1999, \apj, 523, 701

\bibitem[{{Strong} \& {Mattox}(1996)}]{Strong:96}
{Strong}, A.~W. \& {Mattox}, J.~R. 1996, \aap, 308, L21

\bibitem[{{van Dokkum}(2008)}]{VanDokkum:08}
{van Dokkum}, P.~G. 2008, \apj, 674, 29

\bibitem[{{Wilkins} {et~al.}(2008){Wilkins}, {Hopkins}, {Trentham}, \&
  {Tojeiro}}]{Wilkins:IMF}
{Wilkins}, S.~M., {Hopkins}, A.~M., {Trentham}, N., \& {Tojeiro}, R. 2008,
  ArXiv e-prints

\end{thebibliography}

\end{document}